\documentclass{aa}

\usepackage{lineno}
\usepackage{float}
\usepackage{graphicx}
\usepackage{txfonts}

\usepackage{listings}
\usepackage{gensymb}
\usepackage{amsmath,amsfonts,amssymb}
\usepackage{pifont}
\usepackage{isomath}
\usepackage{xcolor}

\newcommand{\pz}[1]{#1}

\begin{document}

   \title{Location and energy of electrons producing the radio bursts from AD Leo observed by FAST in December 2021}


   \author{Philippe Zarka\inst{1,2}, Corentin K. Louis\inst{1},
           Jiale Zhang\inst{3,4,5}, \\ Hui Tian\inst{3}, 
           Julien Morin\inst{6} \and Yang Gao\inst{7}
          }
   \institute{LESIA, Observatoire de Paris, CNRS, Universit\'{e} PSL, Sorbonne Univ., Univ. Paris Cit\'{e}, F-92190 Meudon, France\\   \email{philippe.zarka@obspm.fr}
    \and ORN, Observatoire Radioastronomique de Nan\c{c}ay, Observatoire de Paris, CNRS, Univ. PSL, Univ. Orl\'{e}ans, F-18330 Nan\c{c}ay, France 
    \and School of Earth and Space Sciences, Peking University, Beijing 100871, People's Republic of China
    \and ASTRON, Netherlands Institute for Radio Astronomy, Oude Hoogeveensedĳk 4, Dwingeloo, 7991 PD, The Netherlands
    \and Kapteyn Astronomical Institute, University of Groningen, P.O. Box 800, 9700 AV, Groningen, The Netherlands
    \and Laboratoire Univers et Particules de Montpellier, Universit\'{e} de Montpellier, CNRS, 34095, Montpellier, France
    \and School of Physics and Astronomy, Sun Yat-Sen University, Zhuhai, 519082, Guangdong, China
    }



   \date{Received ?; accepted ?}

 
  \abstract
   {In a recent paper, we presented circularly polarized radio bursts detected by the radio telescope FAST from the flare star AD Leo on December 2-3, 2021, which were attributed to the electron cyclotron maser instability.}
   {In that context we use here two independent and complementary approaches\pz{, inspired from the study of auroral radio emissions from solar system planets,} to constrain for the first time the source location (magnetic shell, height) and the energy of the emitting electrons.}
   {These two approaches consist of (i) modeling the overall occurrence of the emission with the ExPRES code, and (ii) fitting the drift-rate of the fine structures observed by FAST.}
   {We obtain consistent results pointing at 20-30 keV electrons on magnetic shells with apex at 2-10 stellar radii. Emission polarization observed by FAST and magnetic topology of AD Leo favour X-mode emission from the southern magnetic hemisphere, from which we draw constraints on the plasma density scale height in the star's atmosphere.}
   {\pz{We demonstrate that sensitive radio observations with high time-frequency resolutions, coupled to modelling tools such as ExPRES, analytical calculations and stellar magnetic field measurements, now allow us to remotely probe stellar radio environments.} We provide elements of comparison with solar system radio bursts (Jovian and Solar), emit hypotheses about the driver of AD Leo's radio bursts and discuss the perspectives of future observations\pz{, in particular} at very low frequencies ($<$100 MHz).}

   \keywords{Radio bursts -- Flare stars -- Stellar flares -- Stellar  magnetic fields -- Radiation processes}

   \keywords{Stars: individual:AD Leonis -- Stars: magnetic field -- Stars: atmospheres -- Radio continuum: stars -- Radio: bursts -- Radiation mechanisms: non-thermal}

   \titlerunning{Location and energy of electrons producing AD Leo's radio bursts}
   \authorrunning{P. Zarka et al.} 
\maketitle

%

\section{Introduction}
\label{sec:intro}

AD Leo is an extensively studied flare star. 
\pz{Early radio observations focused on describing burst morphology, measuring flux density and polarization, putting constraints on source size as well as on the radiation mechanism (plasma emission versus electron cyclotron-maser -- ECM) and its fundamental or harmonic nature \citep[see e.g.,][]{Lang1983,Lang1986,Gudel1989,Abada1994,Abada1997}.}
Using the 305 m telescope of the Arecibo Observatory in L-band, \citet{Osten2006,Osten2008} detected circularly polarized radio bursts at minutes to sub-second time scales, including fast-drifting bursts in the time-frequency (t-f) plane\pz{, and statistically analyzed burst durations, bandwidths and drift-rates, comparing them to those of Solar spikes}. 

Using the Five-hundred-meter Aperture Spherical radio Telescope (FAST), \citet{Zhang2023} $-$ hereafter paper \#1 $-$ detected in December 2021 and characterized these fast-drifting features with exquisite sensitivity and t-f resolutions.
Shortly before that, in late 2020, a Zeeman Doppler imaging campaign using SPIRou characterized the large-scale magnetic field of AD Leo \citep{Bellotti2023}, providing a reasonably good description of this magnetic field at an epoch close to the radio observations with FAST. 
The observed strong radio intensity, large circular polarization degree and fine structures at a time scale of milliseconds led \citet{Zhang2023} to the logical conclusion that the electron cyclotron-maser (ECM) process at the fundamental of the local cyclotron frequency is the most likely generation mechanism for this radio emission. 
We are thus in a favourable context for applying theoretical tools developed primarily for interpreting Jupiter's ECM radio bursts, namely the ExPRES simulation code \citep{Louis2019} and the analysis of burst drift-rates \citep{Zarka1996,Mauduit2023}, in order to derive \pz{for the first time strong} constraints on the locus of the radio source in AD Leo's environment and on the energy of radio-emitting electrons, as well as on the plasma density profile in the star’s atmosphere. 
This is the purpose of the present paper.

\pz{The quantitative remote sensing of stellar magneto-plasma environments is important to distinguish small-scale flaring activity \citep{Aschwanden2006} from large-scale auroral-like dynamics \citep{Hallinan2015}, and to identify the primary engine of the latter (corotation breakdown of ejected plasma or star-planet interactions \citep[see the review by][]{Callingham2024}). Eventually we aim at being able to extend space weather and electron acceleration studies \citep[see e.g.,][]{Prange2004,Morosan2019a,Morosan2019b,Klein2024} to stellar environments, based on radio observations.}

In Section \ref{sec:FAST_observation}, we summarize the main results obtained in paper \#1 from FAST observations of December 2021. In Section \ref{sec:AD_Leo_MFL_chara}, we recall the physical characteristics of AD Leo and especially its magnetic field as deduced from SPIRou's measurements of late 2020. In Section \ref{sec:ExPRES_analysis}, we present the analysis of the radio emission envelope at a time scale of minutes, using the ExPRES code. In Section \ref{sec:analytical_study}, we present our analysis of the burst drift-rates measured by FAST. We summarize and discuss our conclusions in Section \ref{sec:discussion_conclusions} and give some perspectives for further work in Section \ref{sec:perspectives}.

\section{Summary of FAST observation of AD Leo}
\label{sec:FAST_observation}
As a large single dish -- 300 m instantaneous diameter within a 500 m mirror --, FAST \citep{Nan2011,Jiang2020} has a modest angular resolution (of order of 3') and thus suffers large confusion noise, but it is well adapted to emissions varying with time and/or frequency, for which it provides high instantaneous sensitivity (a few mJy).
FAST observations were performed in L-band (i.e. in the range $\sim$1000--1500~MHz), with sampling time $\sim$0.2 msec in 1024 frequency channels (0.5 MHz spectral resolution), with full-polarization (4 Stokes).
AD Leo observations reported in paper \#1 were performed on December $2^{nd}$ and $3^{rd}$, 2021, for 3 hours each, from 20:30 to 23:30 UT. Only data in the two clean bands 1004--1146~MHz and 1293--1464~MHz were analyzed in paper \#1, the rest being partly polluted by radio frequency interference.

\begin{figure}[!ht]
	\begin{center}
	\centerline{\includegraphics[width=1.\linewidth]{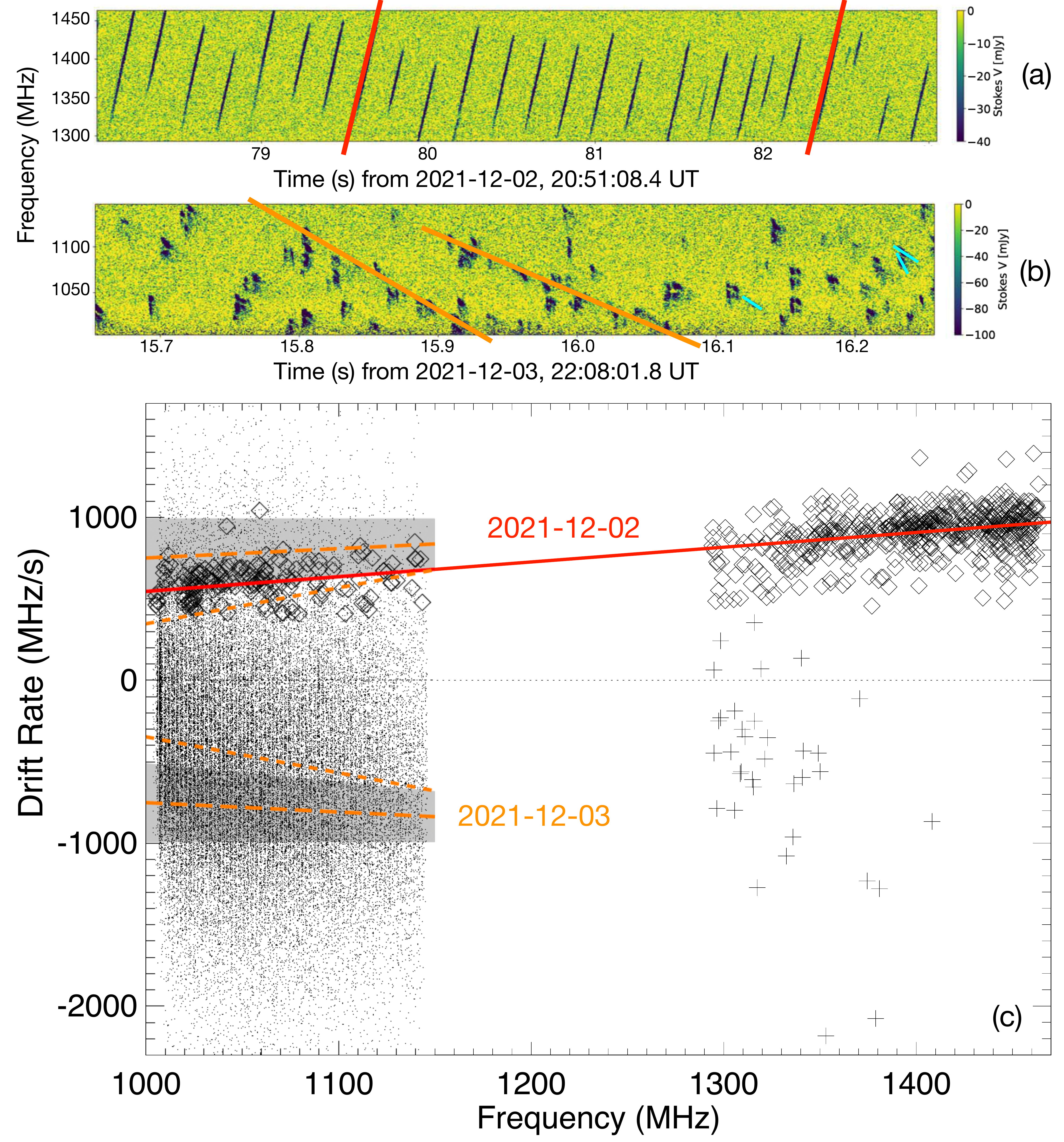}}
\caption{Bursts and drift-rates ($df/dt$) observed with FAST. (a) Representative examples of bursts observed on December $2^{nd}$. Many linearly drifting discrete bursts show up clearly. About 700 individual bursts were identified and their drift-rate measured across the observed frequency range (examples are displayed in red). With FAST, negative Stokes V correspond to LH circular polarization \citep{Wang2023}. (b) Representative examples of bursts observed on December $3^{rd}$. Their morphology is quite different from the previous day. About 50000 individual sub-bursts were identified and their instantaneous drift-rate estimated (examples are displayed in light blue). Examples of overall drifts of sub-burst alignments are displayed in orange. (c) Individual burst drifts on December $2^{nd}$ are displayed as diamonds, and their linear dependence versus the frequency is the best fit red line (similar to paper \#1). Individual sub-burst drifts on December $3^{rd}$ are displayed as dots below 1150 MHz and `+' above 1290 MHz. Their distribution is much more dispersed than the one on the previous day. The best fit linear trend of $df/dt(f)$ for individual sub-burst is the short-dashed orange line between -350 and -680 MHz. Overall drifts of sub-burst trains or clusters are largely spread between -500 and -1000 MHz/s (lower grey-shaded region) and display a small statistical variation between about -750 MHz at 1000 MHz and -840 MHz at 1150 MHz (long-dashed orange line). Determination of these overall drifts is detailed in Appendix \ref{appendix:overall_drift_rate}. The upper grey-shaded region and dashed orange lines are the symmetrical of the lower ones with respect to the $df/dt=0$ dotted line.}
\label{drifts_obs}
\end{center}
\end{figure}

\begin{figure*}[!ht]
	\begin{center}
	\centerline{\includegraphics[width=1.0\linewidth]{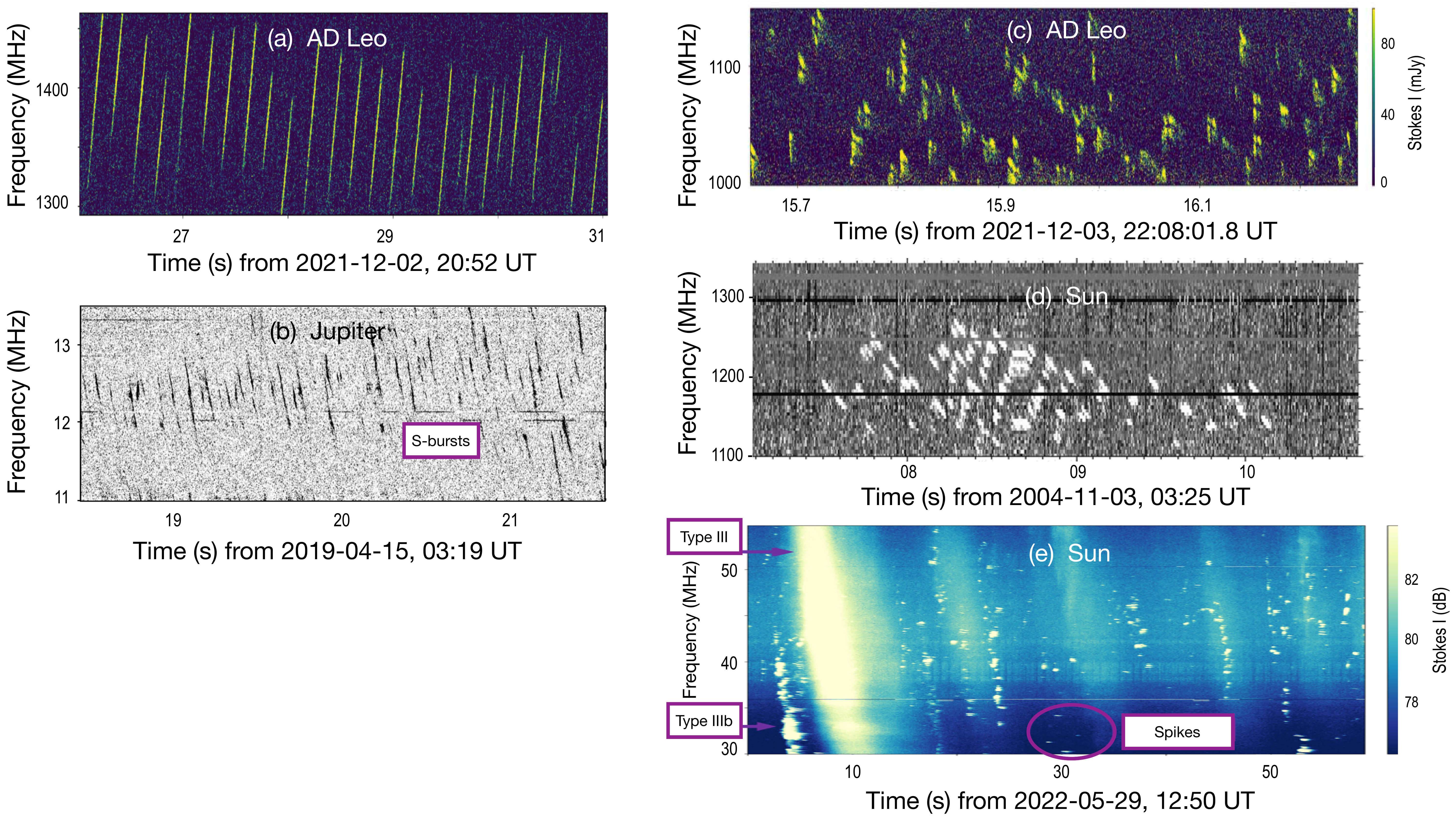}}
\caption{Comparisons of the morphologies of AD Leo's radio bursts observed on December $2^{nd}$, 2021 (a) and December $3^{rd}$, 2021 (c), with typical Jupiter S-bursts observed at the Nan\c{c}ay Decameter Array \citep{lamy2017,Mauduit2023} (b), and with Solar decameter spike bursts observed with the Huairou/NAOC solar spectrometer \citep{Chernov2008} (d) and NenuFAR low-frequency radio telescope \citep{Zarka2020,Briand2022} (e).}
\label{comparisons}
\end{center}
\end{figure*}

On December $2^{nd}$, intense emission including bursts was detected for 8 minutes (20:45–20:53 UT), over the entire frequency range 1000-1470 MHz.
Linearly drifting bursts were detected with positive drift-rates from +550 MHz/s at 1000 MHz to +970 MHz/s at 1470 MHz, with dispersion of $\pm$200 MHz/s around this general trend (cf. Fig. 3a of paper \#1 and Figures \ref{drifts_obs}a,c of the present paper). 
Individual bursts had a typical instantaneous bandwidth $\sim$3.5 MHz and fixed-frequency duration $\sim$6 msec, and drifted over 50 to 100 MHz in about 100 msec (Fig. \ref{drifts_obs}a).
Flux density of the bursts reached 188 mJy (average $\sim$100 mJy).
The emission was strongly Left-Hand (LH) circularly polarized\footnote{RH circular polarization was mentioned in paper \#1, but it has been corrected after comparison with pulsars with known circular polarization and after checking the definition of the polarization measured with FAST \citep{Wang2023}. A erratum has been published \citep{Zhang2024}.}, with an average degree V/I $\sim$35\% depending on the frequency (from $\sim$20\% at 1000 MHz to $\sim$40\% at 1470 MHz), the instantaneous circular polarization degree of bursts reaching 100\% at times. The morphology of these bursts is strikingly similar to that of Jupiter's S-bursts detected at lower frequencies, except for the sign of the drift-rate, as illustrated in Figure \ref{comparisons}a,b \citep[cf. also Figs. 1 of][]{Hess2009,Ryabov2014}. Together with the measured circular polarization, it strongly suggests that these bursts are generated by the ECM mechanism.

On December $3^{rd}$, intense emission including bursts was detected for 90 minutes (21:13–22:48 UT), over the frequency range 1000-1150 MHz only (except for an interval shorter than 1 s reaching 1400 MHz).
Embedded bursts displayed a complex structure at various time scales, consisting of slightly elongated blobs and spots -- hereafter called sub-bursts -- of typical individual duration 2-15 msec and bandwidth $\sim$2.5 MHz (Fig. \ref{drifts_obs}b). 
Sub-bursts were sometimes gathered in pairs separated by a few MHz and in 100--200 msec-long trains or clusters aligned with an overall drift-rate between -500 and -1000 MHz/s (lower grey-shaded region of Figure \ref{drifts_obs}c). Average drift-rates of sub-burst trains, further discussed in Appendix \ref{appendix:overall_drift_rate}, statistically vary from about -750 MHz at 1000 MHz to -840 MHz at 1150 MHz (long-dashed orange line of Fig. \ref{drifts_obs}c).
Flux density of the bursts reached 680 mJy (average $\sim$140 mJy).
The emission was again strongly LH circularly polarized$^1$, with an average degree V/I $\sim$45\%.
Largely overlapping with and prolongating the FAST observation, the upgraded GMRT (uGMRT) detected AD Leo's burst of December $3^{rd}$ in the range 550–850 MHz with a coarser time resolution of 5 s \citep{Mohan2024}. The observation started at 21:17:41 UT and lasted for 7 hours. LH polarized emission was detected across the entire uGMRT band simultaneous to the emission detected by FAST, with a morphology reminiscent of a group of solar Type III bursts \citep[cf. Fig. 1 of][]{Mohan2024}. And indeed, 
the morphology of the bursts detected by FAST is reminiscent of that of some solar spike bursts accompanying type III bursts \citep{Chernov2008}, as displayed in Figure \ref{comparisons}c,d. At much lower frequencies, around 50 MHz, NenuFAR \citep{Zarka2020} observations also sometimes display polarized spiky blobs accompanying Type III bursts (cf. Figure \ref{comparisons}e).
After the end of FAST observation at 23:30, another emission was detected by the uGMRT above 700 MHz and interpreted as a solar Type IV burst \citep{Mohan2024}.
Despite morphological differences, bursts on December $2^{nd}$ and sub-bursts on December $3^{rd}$ display similar fixed-frequency duration and instantaneous bandwidth, while the overall duration of bursts on December $2^{nd}$ is close to that of sub-burst trains on December $3^{rd}$. As ECM is a popular explanation for solar spikes \citep[see e.g.,][]{Benz1986}, we will also assume below that this mechanism is responsible for the polarized drifting sub-bursts and trains detected by FAST on December $3^{rd}$.

\section{Physical characteristics and magnetic field of AD Leo}
\label{sec:AD_Leo_MFL_chara}

AD Leo (GJ 388) is an M3.5V dwarf with a mass of 0.42 $M_{\odot}$ and radius of $R_* \sim 0.4 R_{\odot}$ \citep{Mann2015}, located at a distance of 4.9651$\pm$0.0007 pc from the solar system \citep{GaiaColl2021}. 
It has a rotation period $P_{rot}=2.2300\pm0.0001$ days \citep{Carmona2023} with an origin ($\Phi=0^\circ$ and longitude $\Lambda=0^\circ$ for the direction of the observer) at HJD=2458588.7573, i.e. its rotational phase at any HJD can be computed as $\Phi(^\circ)$=(HJD-2458588.7573)$\times 360/P_{rot} \mod 360^\circ$ (and $\Lambda=360^\circ-\Phi$). AD Leo's inclination, i.e. angle between its rotation axis and the direction of Earth, is $i=20^\circ$, implying an almost pole-on view \citep{Morin2008}. 

AD Leo is an extensively studied flare star, with an intense magnetic activity revealed by frequent flares \citep{Gudel2003,Hawley2003,vandenBesselaar2003,Robrade2005,Muheki2020,Namekata2020}. 

Previous Zeemann–Doppler imaging (ZDI) studies have suggested that AD Leo has a predominant dipole magnetic field and a nearly pole-on geometry, the magnetic south pole being the visible pole for a terrestrial observer, with magnetic field lines entering into the star \citep{Morin2008,Lavail2018}. Recent ZDI measurements by SPIRou in late 2020 have revealed to be complicated to interpret: the dipolar component still dominates ($\sim$70\%) but with significant contributions of higher order terms, especially the large quadrupolar component ($\sim$21\%), and large residuals of the ZDI modelling of Stokes V profiles \citep{Bellotti2023}. Thus, for purposes such as radio emission modelling, it is considered adequate to use in a first step the purely dipolar fit of AD Leo's magnetic field, where the field amplitude at the magnetic south pole is 923$\pm$70 G, and the dipole misalignment with respect to the rotation axis is $59^\circ \pm 2^\circ$ \citep[Table 1]{Bellotti2023}. 
Note however that small-scale field structure likely exists \citep{Yadav2015,Bellotti2023}, although it should rapidly decrease with the altitude above the photosphere. Long term evolution of AD Leo's magnetic field also exists, as demonstrated by \citet{Bellotti2023}, which justifies the use of the field description based on ZDI measurements as close as possible in time to FAST observations for our radio emission modelling \citep[configuration 2020b from][]{Bellotti2023}.

Finally, let us recall that the analysis of radial velocity measurements led \citet{Tuomi2018} to suggest the existence of a giant planet with a mass of $\sim 0.24 M_{Jup}$ in spin–orbit resonance (orbital period of 2.23 days), but that was refuted by subsequent studies which attributed radial velocity variations to the stellar activity \citep{Carleo2020,Carmona2023}. A recent study ruled out planets more massive than 27 $M_\oplus$ orbiting at the stellar rotation period, as well as planets more massive than $3-6 M_{Jup}$ with periods up to 14 years \citep{Kossakowski2022}.

\section{ExPRES analysis of radio burst envelopes}
\label{sec:ExPRES_analysis}

Assuming that the radio bursts from AD Leo detected by FAST are generated via the ECM mechanism at the fundamental of the local cyclotron frequency, we can use the ExPRES code \citep[Exoplanetary and Planetary Radio Emissions Simulator --][]{Louis2019} to derive constraints on the location and energy of the electrons producing the radio emission. ExPRES was developed for simulating the dynamic spectra of Jupiter's decameter radio emissions \citep{Hess2008}, and more precisely the t-f occurrence and the sense of circular polarization of the emissions (not their intensity nor polarization degree). Inputs to the code include the type of electron distribution driving the ECM (loss-cone or shell in the velocity space), the characteristic energy of the electrons (in the case of a loss-cone), a magnetic field model at the source (i.e. of Jupiter or, in our case, of AD Leo), the location of the radio sources (e.g. along field lines with fixed longitude that rotate with the planet or star), the thickness of the hollow conical beam produced by the ECM (usually $1^\circ-2^\circ$) and the position of the observer (fixed or moving). The code then populates the source field lines with radio sources at the local electron cyclotron frequency ($f_{ce} = eB/2 \pi m$, with $B$ the amplitude of the local magnetic field and $e$ and $m$ the charge and mass of the electron), computes the radio beaming angle at each frequency, i.e. the angle relative to the local magnetic field at which the radio emission is beamed (that depends on the frequency and on the electrons energy for loss-cone-driven ECM $-$ it is always $90^\circ$ for a shell-driven ECM), and compares it to the direction of the observer at each time step. When the difference is less than the beam thickness, the emission is considered detected at the corresponding time and frequency. Emissions produced from a northern magnetic hemisphere are RH circular if on the extraordinary mode (so-called R-X mode), and (Left-Hand) LH if on the ordinary mode (L-O mode). Opposed senses of polarization are emitted from a southern magnetic hemisphere. Near-source refraction can be taken into account if a plasma model is available. 

In the original ExPRES paper \citep{Louis2019}, the expression of the refraction index and thus of the radio beaming angle were presented in a condensed form and for the R-X mode only. Appendix \ref{appendix:dispersion_equation_solution}
of the present paper provides their complete derivation for both R-X and L-O modes \citep[ExPRES Version 1.3.0,][]{louis_2023_expresV130}.

Following \citet{Bellotti2023} we have considered for AD Leo a magnetic dipole of moment 461.5 G.R$_*^3$ (i.e. an equatorial surface field of 461.5 G), inclined at 59$^\circ$ from the rotation axis, itself at $\sim 20^\circ$ from the line of sight, with the magnetic south pole in the hemisphere visible for a terrestrial observer. The star rotates in $2.23$~days according to the phase system defined in Section \ref{sec:AD_Leo_MFL_chara}. The magnetic dipole is assumed to be centered on the star's center, and we have not considered the star flattening (negligible for our simulations). AD Leo being a relatively cool star, its coronal plasma density likely drops rapidly above the photosphere and the ambient plasma frequency is likely smaller than the frequency of FAST observations ($\geq 1$~GHz). As a consequence, we neglect near-source refraction and thus we assume straight line propagation from the radio source to the terrestrial observer. This is further discussed in Section~\ref{sec:discussion_conclusions}.

\begin{figure}[!ht]
	\begin{center}
	\centerline{\includegraphics[width=1.0\linewidth]{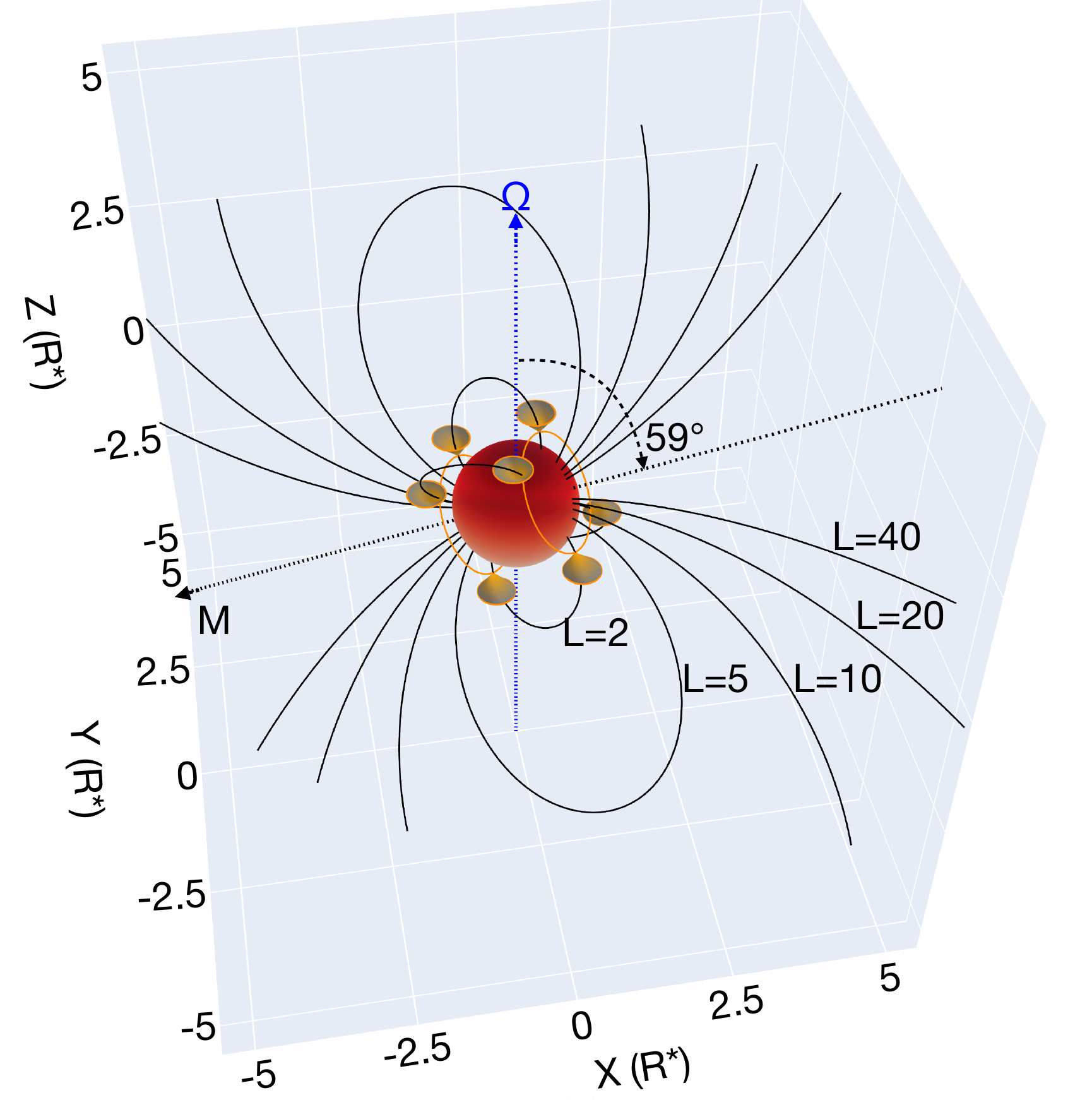}}
\caption{Sketch of AD Leo's magnetic configuration, based on \citep{Bellotti2023}. The axes (x, y, z) are expressed in stellar radius $R_*$. The dotted blue line labelled $\Omega$ represents the star's rotation axis (it is displayed at 20$^\circ$ from the line of sight), while the black dotted line labelled $M$ represents the magnetic dipole axis, making an angle of 59$^\circ$ with $\Omega$. Magnetic field lines of L-shell 2 to 40 are displayed as the solid black lines, for longitudes 20$^\circ$ and 200$^\circ$ (field lines at longitudes 110$^\circ$ and 290$^\circ$ are added for L=2). The auroral ovals (actually circles) at L=2 and $f_{ce}$=1 GHz in both magnetic hemispheres are displayed in orange. Examples of hollow emission cones are shown at the intersection of the ovals and the displayed magnetic field lines at L=2. \pz{Cone apices have a brighter orange shading. Cone wall thickness is figured by the orange circles along cone edges. Cones pointing upward (i.e. toward the observer's hemisphere) can be identified by the field lines visible inside them, whereas cones pointing downward (away from the observer) mask the field lines which carries them. }In our simulations, we place the radio sources and emission cones at every degree of longitude.}
\label{sketch}
\end{center}
\end{figure}

Figure \ref{sketch} is a sketch of the geometry of AD Leo's dipolar magnetic field as seen from a terrestrial observer, with a few radio emission cones displayed.

\begin{figure}[!ht]
	\begin{center}
	\centerline{\includegraphics[width=1.0\linewidth]{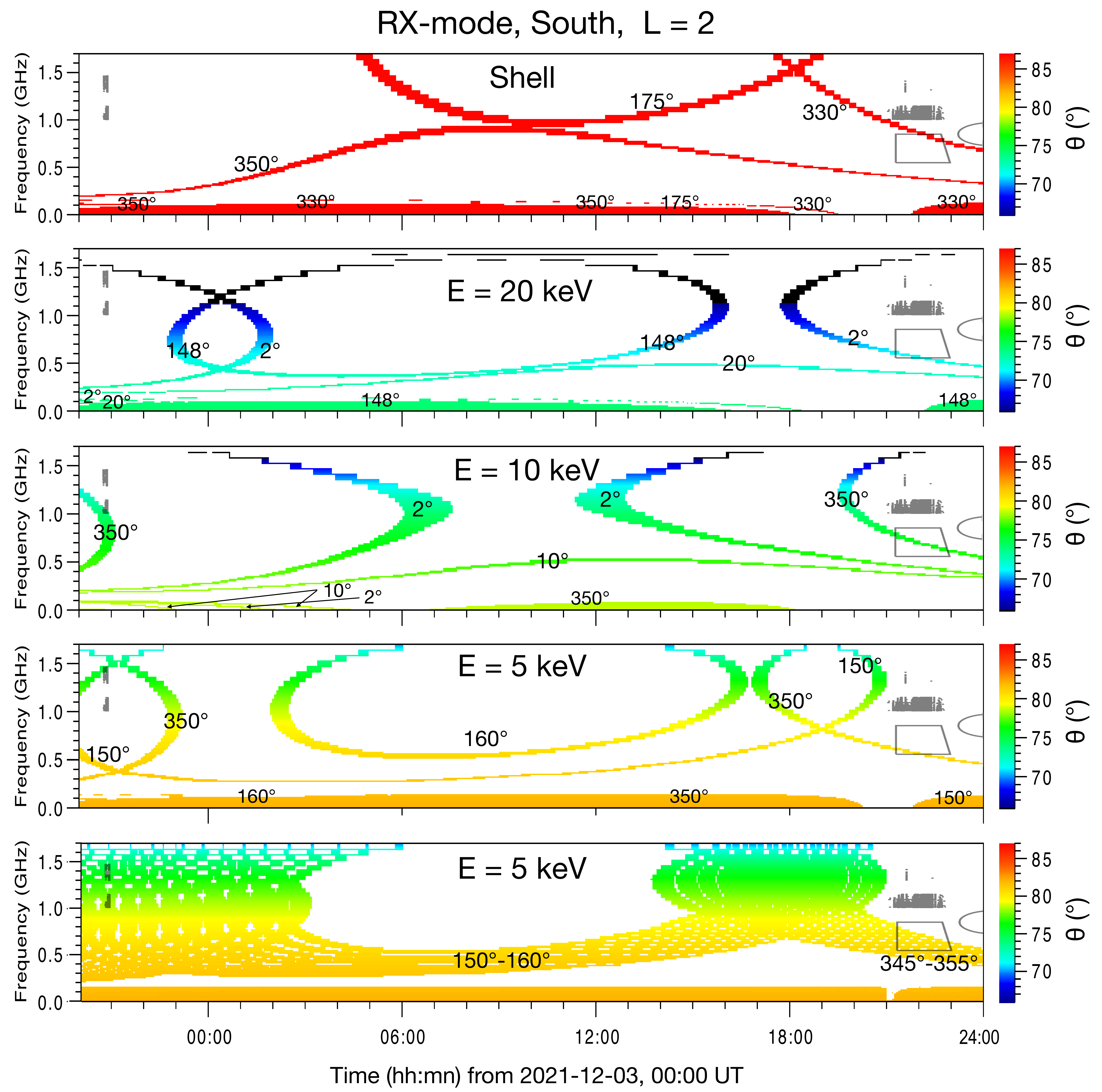}}
\caption{Sample ExPRES simulations of ECM radio emissions from AD Leo between Dec. $2^{nd}$, 20:00 UT and Dec. $3^{rd}$, 24:00 UT. Emitting radio sources are placed along AD Leo's dipolar field lines with magnetic L-shell=2 in the southern hemisphere (in view of the observer), at longitudes indicated on the figure. Emission is produced permanently along each entire field line at the local $f_{ce}$ on the R-X mode, and beamed in a hollow cone of aperture self-consistently computed by ExPRES (see Appendix \ref{appendix:dispersion_equation_solution}) and of thickness $1^\circ$. 
\pz{The upper 4 panels display simulated emissions from a single field line at each indicated longitude. The lower panel displays simulated emissions from 2 sets of field lines covering each a $10^\circ$ longitude range with one field line per degree of longitude (the small t-f gaps, that result from this discretization, disappear with a denser filling, i.e. with more field lines per degree of longitude).}
Radio arcs are observed across the t--f plane from 0 to 1.7 GHz. Signals observed below $\sim$0.2 GHz result from a mix of emissions produced at various longitudes. Each panel explores one emission scenario for the ECM driver (loss-cone with 5, 10 or 20 keV characteristic energy, or shell in the velocity space). The color scale indicates the beaming angle $\theta (^\circ)$ of the emission relative to the local magnetic field vector. Bursts were detected by FAST in the grey-shaded areas, while the grey contours refer to uGMRT detections.}
\label{expres}
\end{center}
\end{figure}

In Figure \ref{expres}, we show typical outputs from ExPRES applied to AD Leo, in which a few selected field lines (of apex distance at magnetic equator 2 $R_*$, i.e. shell parameter L=2, and of longitudes $2^\circ$, $10^\circ$, $20^\circ$, $148^\circ$, $150^\circ$, $160^\circ$, $175^\circ$, $180^\circ$, $270^\circ$, $330^\circ$ and $350^\circ$\pz{, as well as the ranges of longitudes $150^\circ-160^\circ$ and $345^\circ-355^\circ$ with one field line per degree of longitude}) are assumed to radiate at the local $f_{ce}$ on the R-X mode from the southern hemisphere. The simulations are performed over 28 hours across December $2^{nd}$ and $3^{rd}$, during which the radio sources rotate with the star. The hollow conical beam thickness is taken equal to $1^\circ$. Four scenarii are tested, in which the ECM is driven by a loss-cone with characteristic energy 5, 10 or 20 keV, or a shell electron distribution. Arcs are detected by a terrestrial observer in the t-f plane in the frequency range covered by FAST.
The envelopes of the radio bursts detected on December $2^{nd}$ and $3^{rd}$ are displayed as grey-shaded areas at about -03:00 (i.e. $\sim$21:00 on December $2^{nd}$) and 21:30 to 23:00. For reference, uGMRT detections are indicated as the grey contours.
We note that the ExPRES arcs \pz{from a single field line} displayed in \pz{the upper 4 panels of} Figure \ref{expres} have fixed-frequency durations comparable to the envelopes of the observed bursts\pz{, whereas emission from a set of field lines spreading over $10^\circ$ of longitude produce much more extended regions in the t-f plane (lower panel of Figure \ref{expres})}. This suggests that only a restricted range of stellar field lines were emitting in radio at the time of FAST observations, not necessarily the same on the two days.

In order to limit the number of free parameters and make the minimum \textit{ad hoc} assumptions on the radio source, the modelling presented below assumes an auroral-like emission from AD Leo, where radio emission can be produced at all longitudes. The set of radio-emitting field lines is thus characterized only by its magnetic shell parameter L, i.e. the distance of the field line apex to the center of the star. We consider separately R-X and L-O modes from the northern or southern hemisphere of the star. For a given value of L, we place radio sources at every degree of longitude along field lines of parameter L, and radio waves are assumed to be produced permanently along each entire field line, from the surface of the star to the apex of the field line, at the local $f_{ce}$ at each point. The thickness of the hollow cone produced by each point source is taken equal to $1^\circ$, in order to ensure overlapping between the cones produced by consecutive sources separated by $1^\circ$ of longitude, resulting in a quasi-continuous coverage of the parts of the t-f plane where emission is detected. With this modelling, we compute thus a ``maximum'' coverage of the t-f plane for each selected L-shell, emission mode, source hemisphere, and ECM energy source. 

\begin{figure}[!ht]
	\begin{center}
	\centerline{\includegraphics[width=1.0\linewidth]{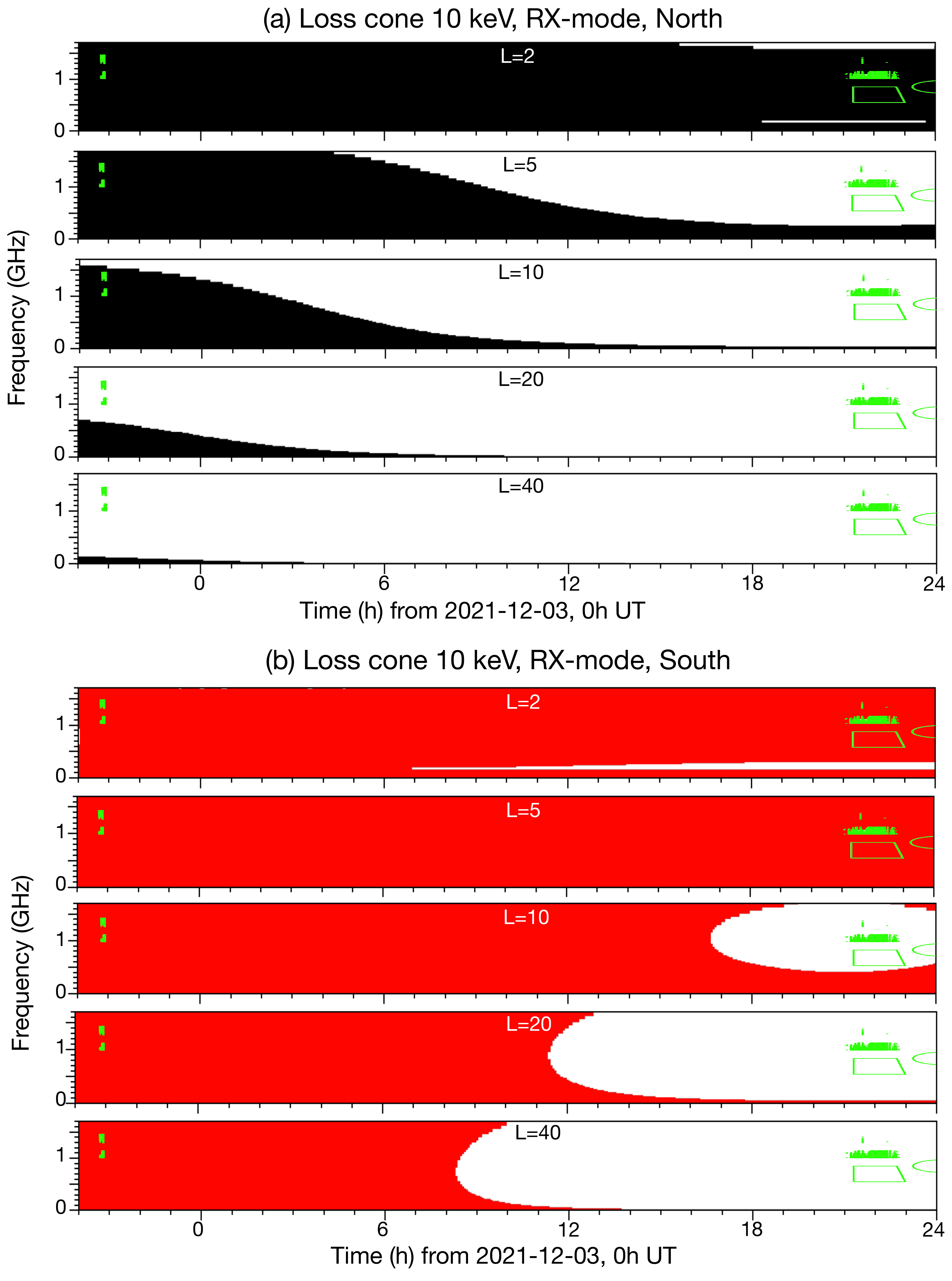}}
\caption{Examples of simulated emission envelopes with ExPRES. R-X mode is emitted from the Northern (a) or Southern (b) hemisphere, by loss-cone-driven ECM with characteristic energy 10 keV. Five dipolar magnetic shells (L=2, 5, 10, 20, 40) are simulated in each case, with active radiosources along all field lines at the corresponding shell (actually every degree of longitude \pz{-- small t-f gaps due to this discretization have been interpolated}). An Earth-based observer detects RH (black) or LH (red) polarized emission depending on the hemisphere of origin and emission mode. On both panels, bursts were detected by FAST in the green-shaded areas, while the green contours refer to uGMRT detections.}
\label{envelopes}
\end{center}
\end{figure}

We have produced simulated dynamic spectra in occurrence and sign of circular polarization for seven ECM energy sources (loss-cone electron distribution with characteristic electron energy of 5, 10, 30, 100, 200 and 500 keV, and shell electron distribution), and five values of the source L-shell (L=2, 5, 10, 20 and 40 stellar radii), for a terrestrial observer, from December $2^{nd}$, 20:00 UT to December $3^{rd}$, 24:00 UT. This time interval encompasses both observations by FAST. R-X and L-O modes were simulated from each hemisphere of AD Leo, separately. The criterion to decide the compatibility of a simulation with the observations is that the envelopes of the radio bursts detected by FAST must be included in the t-f region where the simulation predicts emission to occur.

Figure \ref{envelopes} displays significant examples of our ExPRES simulations compared to FAST observations. 
Panel (a) of Fig. \ref{envelopes} displays the simulated dynamic spectra for R-X mode produced from AD Leo's northern magnetic hemisphere by the ECM driven by a loss-cone with characteristic energy 10 keV, for sources placed at every degree of longitude on L-shells 2, 5, 10, 20 and 40 ($R_*$). The part of the t-f plane in which the emission is visible from a terrestrial observer is filled in solid black color, representing RH circular polarization. The t-f areas in which FAST detected bursts are green-shaded. Green contours refer to uGMRT detections.
Panel (a) shows that the corresponding scenario can account for both detected emissions only for the L=2 magnetic shell. The emission observed on December $2^{nd}$ is compatible with a source at L=2 to L=10, whereas that on December $3^{rd}$ is incompatible with sources at L$>$2. In addition, R-X mode from the northern magnetic pole produces RH circular polarization, opposed to the observed one. 
Panel (b) displays the simulated dynamic spectra for the same emission mode and scenario but for the southern hemisphere. Predicted emission is displayed in solid red, representing LH circular polarization. The cases L=2 and L=5 are compatible with both FAST observations, i.e. the simulated dynamic spectra include both green areas, and the predicted polarization corresponds to the observed one.

Similar ExPRES simulations for other loss-cone energies and for a shell of electrons are displayed in Figure \ref{envelopesA} of Appendix \ref{appendix:expres_simulations}.
For a 30 keV loss-cone, predicted R-X southern emission is compatible with FAST observations for L=2 to 10 (Panel (d) of Fig. \ref{envelopesA}). R-X northern emission at L=2 is marginally compatible with the observations, in the sense that the simulated emission includes the observed t-f ranges except the brief extension to 1400 MHz on December $3^{rd}$ (Panel (c)).
For a 100 keV loss-cone, predicted R-X northern emission is incompatible with the observations for all L-shells (Panel (e) of Fig. \ref{envelopesA}), whereas predicted R-X southern emission at L=5 is compatible with the observations (Panel (f)). 
For ECM driven by a loss-cone with characteristic energy 200 keV, no simulated dynamic spectrum matches both FAST observations together (Panels (g) and (h) of Fig. \ref{envelopesA}). 
Finally, panels (i) and (j) of Figure \ref{envelopesA} display the simulated dynamic spectra for R-X mode produced from AD Leo's northern and southern magnetic hemispheres by the ECM driven by a shell of electrons. In that case, the emission is beamed at 90$^\circ$ from the magnetic field in the source whatever the electron's characteristic energy (which is therefore unconstrained). The figure shows that the case L=2 is consistent with FAST observations. Because of the radio beaming angle at 90$^\circ$, the t-f coverage for shell-driven ECM emission from both hemispheres is identical, but with opposed polarization.

We have also performed all simulations for the L-O mode, and found that the predicted t-f coverage is identical for R-X and L-O modes from the same hemisphere, but with opposed polarizations. \pz{Finally we have checked that doubling the cone thickness ($2^\circ$) marginally changes the t-f coverages in Figure \ref{envelopes} and thus does not impact our results. Halving the cone thickness ($0.5^\circ$) only creates many small gaps in the t-f coverages, such as those in the lower panel  of Figure \ref{expres}.}

Table \ref{table1} summarizes the results of our $7 \times 5 \times 2 \times 2 = 140$ ExPRES simulations compared to FAST observations. The predicted t-f coverages are consistent with the observations for loss-cone-driven ECM with 5-100 keV electrons on field lines with L$\leq$5 to 10, on the R-X or L-O mode from the southern magnetic hemisphere.
They are also consistent with 5-10 keV loss-cone-driven ECM from the northern magnetic hemisphere, and for shell-driven ECM in both hemispheres, but only on field lines with L=2. 
When polarization is taken into account, only R-X mode from the southern hemisphere or L-O mode from the northern hemisphere remain compatible with the observations. The corresponding results are emphasized in boldface style in table \ref{table1}.

\begin{table*}[!ht]
\caption{Comparison of ExPRES simulations with FAST observations: L-shell ranges for which the simulated t-f domain includes the observed bursts.}
\label{table1}
\centering
\begin{tabular}{c c c c c}
\hline
Electron distribution feature driving the ECM & {\bf L-O North}  & L-O South & R-X North & {\bf R-X South} \\
\hline
loss-cone 5 keV & {\bf L=2} & L$\leq$5 & L=2 & {\bf L$\leq$5} \\
loss-cone 10 keV & {\bf L=2} & L$\leq$5 & L=2 & {\bf L$\leq$5} \\
loss-cone 30 keV & {\bf (L=2)} & L$\leq$10 & (L=2) & {\bf L$\leq$10} \\
loss-cone 100 keV & -- & L=5 & -- & {\bf L=5} \\
loss-cone 200 keV & -- & -- & -- & -- \\
loss-cone 500 keV & -- & -- & -- & -- \\
Shell & {\bf L=2} & L=2 & L=2 & {\bf L=2} \\
\hline
\end{tabular}
 \tablefoot{For each scenario, emission mode and hemisphere of origin of the emission, the L-shells for which the simulated t-f domain includes the observed bursts are listed. Values in parentheses correspond to cases where the agreement is marginal (see text). Boldface style indicates compatibility with observed polarization.}
\end{table*}

The height of the radio sources between 1000 and 1500 MHz is between 1.10 and 1.23 $R_*$ (for L=2) and at most between 1.19 and 1.34 $R_*$ (for L=10).
The latitude of the radio sources is between $40^\circ \pm 2^\circ$ (for L=2) and $69^\circ \pm 1^\circ$ (for L=10).

In order to test the dependence of our modelling results on AD Leo's dipolar magnetic field parameters, we have re-computed the plots of Figure \ref{envelopes} with the 2019b model from \citet{Bellotti2023} (magnetic dipole of moment 441 G.R$_*^3$ inclined at 23$^\circ$ from the rotation axis). Predicted t-f domain covered by ECM emission slightly change, without modifying significantly the conclusions of Table \ref{table1}. As recent ZDI observations from AD Leo do not suggest any polarisation reversal since 2020, we consider our modelling of Figure \ref{envelopes} as representative of the situation in late 2021, at the time of FAST observations. 

We have also performed ExPRES simulations where magnetic field lines carry radio-emitting electrons only in a restricted longitude range (such as in Figure \ref{expres}), in order to test the possibility of a radio emission induced by the presence of a planet in synchronous orbit, as proposed by \citet{Tuomi2018}. We did not find any single restricted longitude range that could account for the emissions observed by FAST on both December $2^{nd}$ and $3^{rd}$.
Thus, in the frame of our simulations (ECM mechanism at the fundamental of the local cyclotron frequency), FAST observations on December 2-3, 2021 cannot be attributed to a star-planet interaction with a planet in synchronous orbit. More generally, emission from a restricted sector of stellar longitude that would be active on both days is excluded.

\section{Analytical study of burst drift-rates}
\label{sec:analytical_study}

Fast-drifting bursts provide us with an independent way to estimate the source L-shell and electrons energy. For a dipolar magnetic field, the calculations can be conducted analytically. Following \citet{Zarka1996} and \citet{Mauduit2023}, we can write the magnetic field amplitude at distance $R$ from the star's center and colatitude $\theta$ from the magnetic axis as:
\begin{equation}
    B(R,\theta) = \frac{B_e}{R^3}~(1+3~cos^2 \theta)^{1/2}
    \label{eq1}
\end{equation}
with $B_e$ the equatorial surface field (here 461.5 G) and $R$ in $R_*$, and hence the electron cyclotron frequency $f_{ce}$ can be written:
\begin{equation}
    f(R,\theta) = \frac{f_e}{R^3}~(1+3~cos^2 \theta)^{1/2}
    \label{eq2}
\end{equation}
with $f = f_{ce}$ and $f_e$ the cyclotron frequency at the equator at 1 $R_*$ distance. The drift-rate $df/dt$ produced by electrons of energy E moving along a field line L can be written:
\begin{equation}
    \frac{df}{dt} = \frac{df}{d\theta} \times \frac{d\theta}{ds} \times \frac{ds}{dt}
    \label{eq3}
\end{equation}
with $s$ the curvilinear abscissa. Using the equation of a dipolar field line
\begin{equation}
    R(L,\theta) = L~sin^2 \theta
    \label{eq4}
\end{equation}
and with $ds^2 = dR^2 + R^2 d\theta^2$ and $ds/dt = v_{//}$, we obtain:
\begin{equation}
    \frac{df}{dt} = -\frac{3 f g(\theta)}{L R_*} v_{//}
    \label{eq5}
\end{equation}
with
\begin{equation}
    g(\theta)= \frac{cos\theta}{sin^2\theta}~\frac{(3+5cos^2\theta)}{(1+3 cos^2\theta)^{3/2}}.
    \label{eq6}
\end{equation}

Assuming that there is no distributed accelerating electric potential nor any potential drop along the source field line (i.e. the electrons are in adiabatic motion), the expression of the parallel electron velocity is deduced from the total energy E (keV) with the help of the conservation of the energy of the electron ($v^2 = v_{//}^2 + v_{\bot}^2 = constant$), because the magnetic force does not work, and the first adiabatic invariant of the electron motion in a variable magnetic field amplitude ($v_{\bot}^2/B = constant$). We obtain:   
\begin{equation}
    v_{//} = v (1-\frac{f L^3 sin^2\phi_{e}}{f_e})^{1/2}
    \label{eq7}
\end{equation}
with $\phi_{e}$ the equatorial pitch angle of the electron (i.e. $\phi_{e} = arcsin(v_{\bot}/v)$ at the magnetic equator), and $v=c(1-(\frac{E_o}{E+E_o})^2)^{1/2}$
with $E_o = 511$ keV the electron's energy at rest.
Note that the sign of the drift-rate depends on that of $v_{//}$, which reflects the sense of motion of the electrons along the source field line. Downgoing electrons produce positively drifting radio waves (frequency increases with time) whereas upgoing electrons produce negatively drifting signals. The modulus of the drift-rate does not depend on its sign.

The altitude of the mirror point of an electron, at which $v=v_{\bot}$, depends only on its equatorial pitch angle:
\begin{equation}
    \frac{v_{\bot}^2}{B} = \frac{v^2}{B_{mirror}} = \frac{v_{\bot eq}^2}{B_{eq}} = \frac{v^2 L^3 sin^2\phi_{e}}{B_e} 
    \label{eq8}
\end{equation}
that implies
\begin{equation}
    B_{mirror} = \frac{B_e}{L^3 sin^2\phi_{e}},
    \label{eq9}
\end{equation}
the altitude of the mirror point where $B = B_{mirror}$ being given by the expression of $B(R,\theta)$.
Alternately, the equatorial pitch angle on a field line L is expressed as:
\begin{equation}
    \phi_{e} = arcsin(\frac{B_e}{L^3 B_{mirror}})^{1/2} = arcsin(\frac{f_e}{L^3 f_{mirror}})^{1/2}
    \label{eq10}
\end{equation}
The equatorial pitch angle that corresponds to a mirror point at the stellar surface (i.e. at 1 $R_*$) is
\begin{equation}
    \phi_{e1} = arcsin(\frac{f_e}{L^3 f_{max}})^{1/2}
    \label{eq10a}
\end{equation}
with $f_{max}$ the cyclotron frequency at the footprint ($R = 1 R_*$) of the field line L. Electrons with $\phi_{e} < \phi_{e1}$ precipitate into the star and are lost by collisions, generating a loss-cone in the reflected distribution, while electrons with $\phi_{e} > \phi_{e1}$ have their mirror point above the star's surface.

\begin{figure}[!ht]
	\begin{center}
	\centerline{\includegraphics[width=1.0\linewidth]{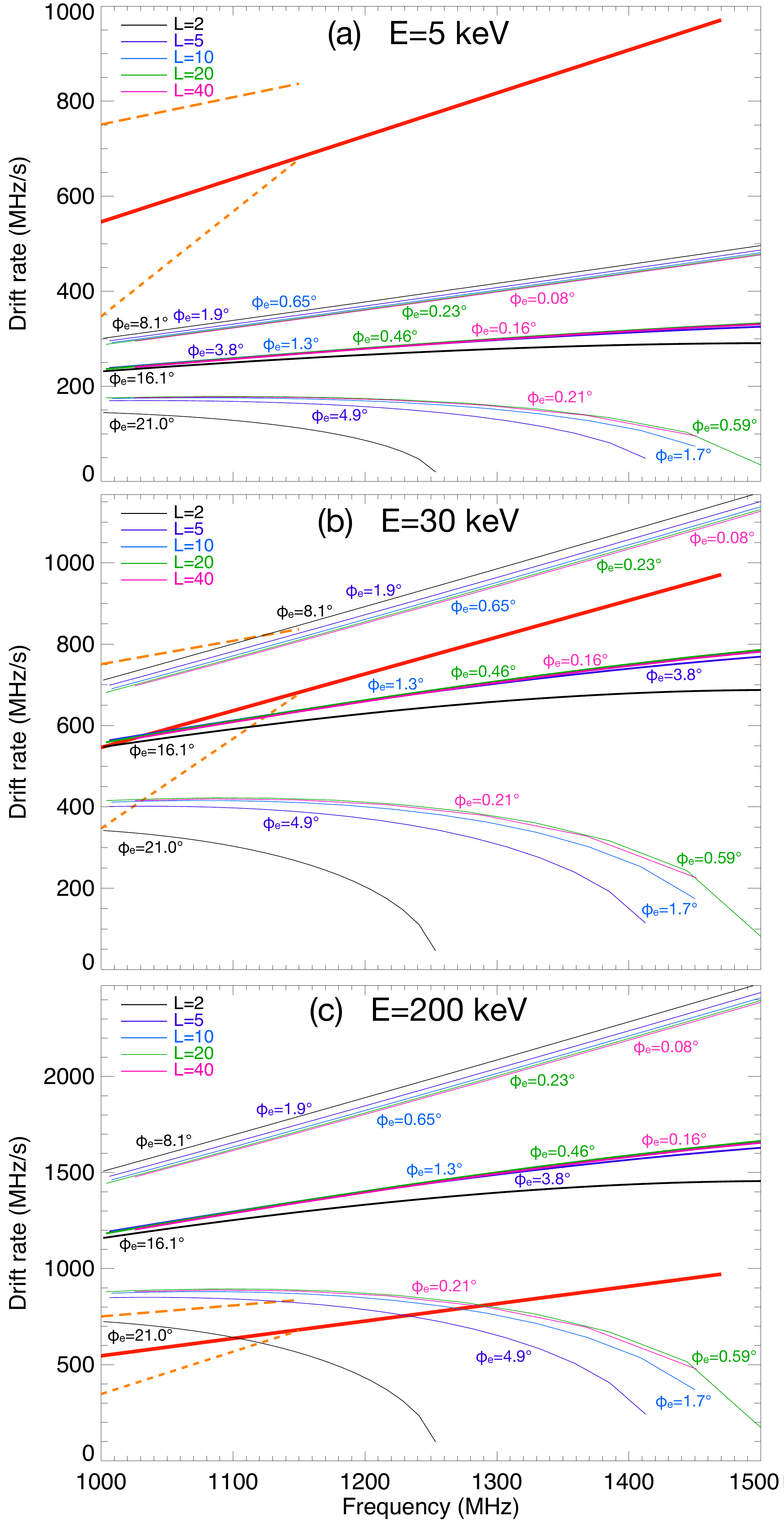}}
\caption{Drift-rates calculated in FAST range for electrons with energy 5 keV (a), 30 keV (b) and 200 keV (c). On each panel, drift-rates for each L value (same values as in Figure \ref{envelopes}) are plotted in a given color (L=2: black, L=5: violet, L=10: blue, L=20: green, L=40: pink). For each L-shell, drift-rates are computed for 3  values of the equatorial pitch angle [$\phi_{e1}/2, ~\phi_{e1}, ~1.3 \times \phi_{e1}$], and the curve corresponding to $\phi_{e} = \phi_{e1}$ is thicker than the other two. The corresponding value of $\phi_{e}$ is indicated next to each curve in the corresponding color. The red and orange lines represent the drift-rates measured by FAST (see text). Increased electron energy corresponds to increased drift-rates proportional to the electron velocity across the source regions.}
\label{drifts_calca}
\end{center}
\end{figure}

\begin{figure}[!ht]
	\begin{center}
	\centerline{\includegraphics[width=1.0\linewidth]{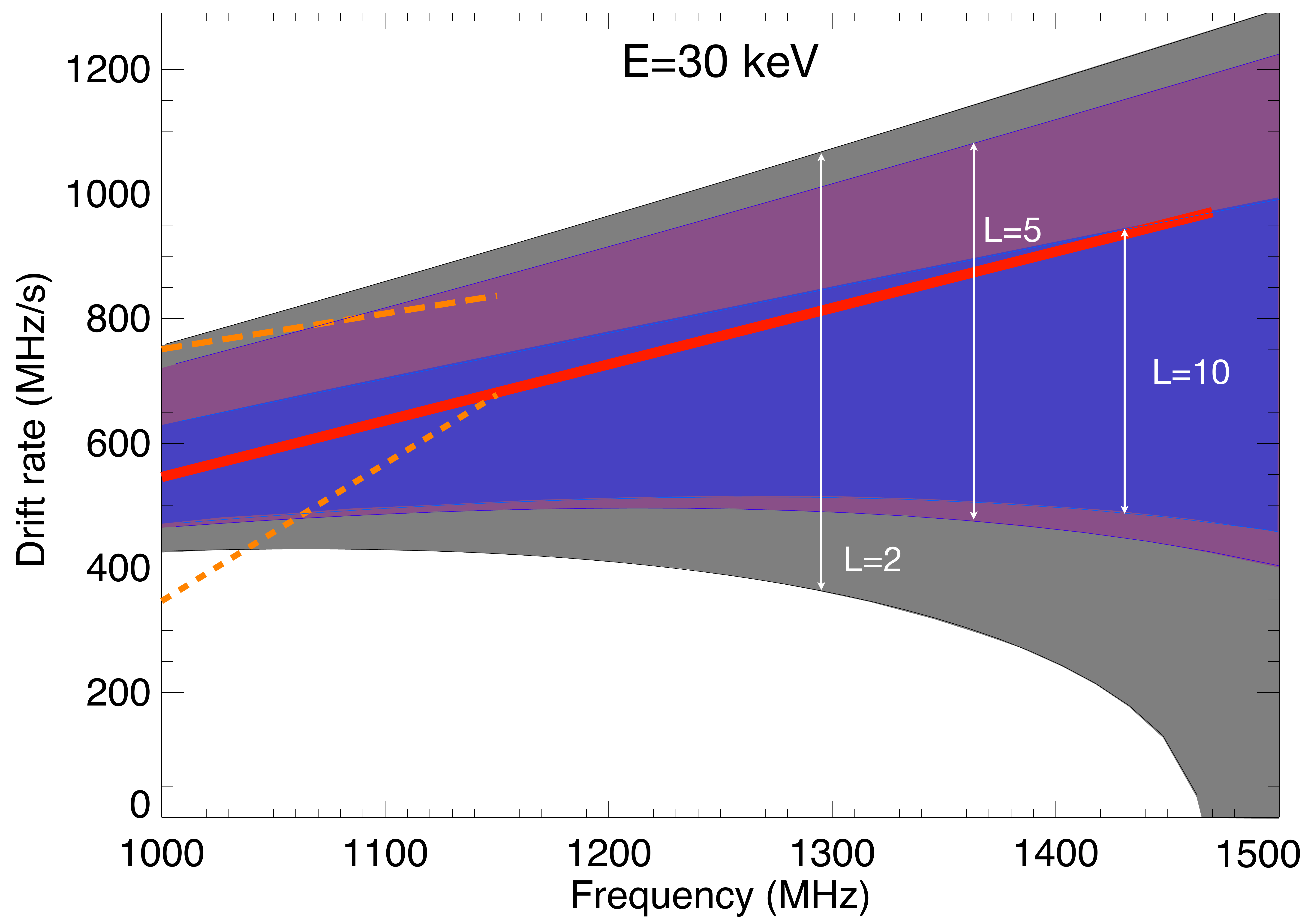}}
\caption{Ranges of calculated drift-rates along dipolar field lines with L-shell 2, 5 and 10, for an electron's energy of 30 keV and equatorial pitch angles between $1^\circ$ (upper limit of each domain) and $1.2 \times \phi_{e1}$ (lower limit). The red and orange lines represent the drift-rates measured by FAST on December $2^{nd}$ and $3^{rd}$ (see text).}
\label{drifts_calcb}
\end{center}
\end{figure}

Using the above equations we have computed predicted drift-rates over the range of FAST observations for electrons with energies of 5 to 200 keV, moving along field lines with L values of 2 to 40.
Figure \ref{drifts_calca} displays representative examples at 5, 30 and 100 keV. The modulus of the drift-rates is plotted for the different L-shells (displayed in different colors) and for 3 values of $\phi_{e}$ on each L-shell (see caption of Figure \ref{drifts_calca}).
The drift-rates measured by FAST on December $2^{nd}$ are displayed with the thick solid red line, while the overall (resp. sub-burst) drift-rates on December $3^{rd}$ are displayed as the long-dashed (resp. short-dashed) orange line, as in Figure \ref{drifts_obs}c. From Figure \ref{drifts_calca} (actually from the entire series of simulations with energies 5-200 keV), we conclude that (i) the drift-rates of December $2^{nd}$ are compatible only with electrons of energy 20-30 keV, (ii) the overall drift-rates of December $3^{rd}$ are compatible only with electrons of energy 30-100 keV, and (iii) the sub-burst drift-rates of December $3^{rd}$ are incompatible with electron adiabatic motion at all energies. 
Moreover, for energies of 20-30 keV, matching the observed drift-rates along field lines with L$>$10 requires electrons moving quasi-purely parallel to the star's dipolar field lines (equatorial pitch angle $\phi_{e} \ll 1^\circ$), which is quite difficult to achieve from low latitude acceleration, that will necessarily lead to an angular dispersion of electron velocities.
If we restrict to the more plausible range $1^\circ \leq \phi_{e} \leq 1.2 \times \phi_{e1}$, we obtain the domains displayed on Figure \ref{drifts_calcb} for an electron energy E=30 keV. Here we see that observed drift-rates can be reached only on field lines with L$\leq$10.

Thus, the results of this analysis are quite convergent with those obtained completely independently from the ExPRES simulations in the previous section, especially for R-X southern emissions. Both are consistent with electrons whose energy is about 30 keV, moving adiabatically along field lines with L$\leq$10. Possible L-O northern emission deduced from ExPRES simulations (column `L-O North' of Table  \ref{table1}) corresponds to electron energies lower than those deduced from drift-rate calculations. Burst drift-rates on December $2^{nd}$ are particularly well matched by 20-30 keV electrons. On December $3^{rd}$, sub-burst drifts are inconsistent with large scale electron motion along dipolar field lines whatever their energy, while overall drift-rates may be consistent with an energy about twice higher than on the previous day.

\section{Discussion and Conclusions}
\label{sec:discussion_conclusions}

As discussed in paper \#1, the high flux density ($>$100 mJy) and brightness temperature ($T_b$ up to $10^{18}$ K) of AD Leo's radio bursts, their high circular polarization degree V/I, and their fine temporal structure at a few milliseconds timescale suggest a generation via the ECM mechanism at the fundamental of the local cyclotron frequency ($f=f_{ce}$), as  for Jovian S-bursts. The latter argument (fine time structure) is especially important, as plasma emission with fine structure much shorter than 1 s is unlikely \citep{Vedantham2021}. Conversely, the very high maser growth rates ensure fast growth possibly up to saturation \citep{Treumann2006}.

\pz{We restricted our study to emission at the fundamental of the cyclotron frequency because the corresponding ECM growth rates are generally much larger than for higher harmonics \citep{Treumann2000,Treumann2006}, so that the fundamental should dominate below $\sim$2 GHz unless it is trapped or absorbed in the plasma. Emission at harmonics of $f_{ce}$ should also reach much higher frequencies than those observed to date by FAST (paper \#1) or by most of the previous observers \citep{Abada1994,Osten2006,Osten2008,Villadsen2019}. Harmonic emission from AD Leo may have been detected in the range 2.8--5 GHz \citep{Stepanov2001,Villadsen2019}.} 

Figures \ref{comparisons}a,b, that compare qualitatively the morphology of AD Leo's bursts from December $2^{nd}$ with Jupiter's S-bursts reveal a striking similitude, the main difference being the sign of the drift-rate that suggests downgoing electrons on AD Leo and upgoing ones at Jupiter. On Figures \ref{comparisons}c,d,e, we also show that the morphology of AD Leo's bursts from December $3^{rd}$ is reminiscent of that of Solar spikes, also commonly attributed to ECM \citep{Benz1986,Wu2007,Chernov2008}. The overall drift of AD Leo's bursts is negative in that case.

The interpretation of FAST observations from December 2021 favours R-X mode from AD Leo's southern magnetic hemisphere, because this mode is consistent with both the t-f coverage and the measured LH polarization. It is also consistent with the fact that the South magnetic hemisphere is in better view from a terrestrial observer.

This has an implication in terms of the density of AD Leo's atmosphere. Fundamental X-mode emission requires $f_{pe}/f_{ce} < 0.3$ (with $f_{pe}$~[kHz]~$\sim~9 N_e^{1/2}$~[cm$^{-3}$] the plasma frequency) in the radio source region \citep{Melrose1984,Treumann2006} in the range of FAST observations, i.e. 1-1.5 GHz. Since AD Leo is a cool red dwarf \citep[$T_{eff} \sim 3500$ K,][]{Mann2015} we can assume in first approximation a corona  in hydrostatic equilibrium with a density (and hence an electron density) varying as $N_e(z) = N_o exp(-z/H)$ (with $z$ the altitude above $r = 1~R_*$). For the Sun, the coronal base density $N_o$ is $N_\odot \sim 3 \times 10^8$ cm$^{-3}$ and the scale height $H$ is $H_\odot \sim 10^8$ m. For a red dwarf, previous authors have assumed a coronal base density $N_* \sim 100 \times N_\odot$ and a scale height $H_* \sim 0.5-1.2 H_\odot$ \citep{Villadsen2019}. Following \citet{Mohan2024} and \citet{Villadsen2019}, we can write the electron density profile in AD Leo's atmosphere as

\begin{equation}
    N_{e}(r)~[cm^{-3}] = n \times 2.5~10^{10} \exp [-(r-1)/(h \times 0.38 R_*)]
    \label{eq11}
\end{equation}

where we have introduced the dimensionless parameters $n$ and $h$ characterizing the coronal base density and scale height relative to the proposed model (i.e. $n=1$ and $h=1$ in \citep{Villadsen2019}). 

\begin{figure}[!ht]
	\begin{center}
	\centerline{\includegraphics[width=1.0\linewidth]{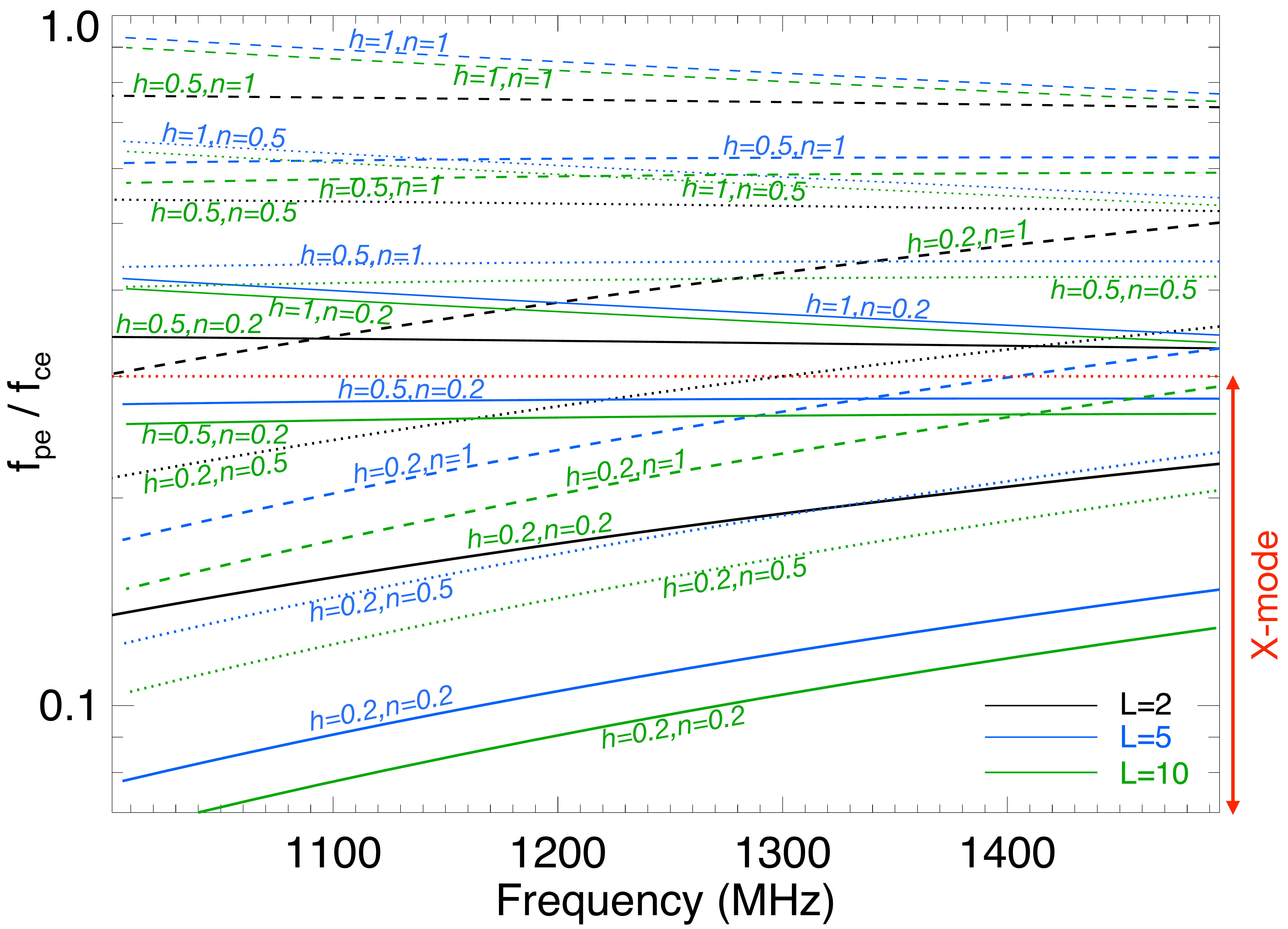}}
\caption{$f_{pe}/f_{ce}$ values in the frequency range of FAST observations, for a dipolar magnetic field of moment 461.5 G.R$_*^3$ and a coronal density following Equation \ref{eq11}, along L-shells 2 and 5, as a function of the relative base coronal density $N$ and relative scale height $H$.}
\label{fpe_fce}
\end{center}
\end{figure}

With the dipole field of AD Leo allowing us to determine $f_{ce}(z)$ along any field line (Equations \ref{eq2} and \ref{eq4}), we obtain on field lines with L-shell 2 to 10 the ratio $f_{pe}/f_{ce}$ displayed on Figure \ref{fpe_fce} for $n \in [0.2, 0.5, 1]$ and $h \in [0.2, 0.5, 1]$. We see on Figure \ref{fpe_fce} that R-X mode sources at L=2 are very unlikely unless both $n$ and $h$ are much smaller than 1. Along field lines with L=5 to 10, fundamental R-X mode generation via ECM in the range 1-1.5 GHz imposes $H \sim 0.2$ (whatever $N$), or $H$ up to 0.5 if $N \sim 0.2$. These parameters correspond to a corona significantly less dense than in \citet{Villadsen2019} at the radio sources altitude (1.10 to 1.34 $R_*$ and latitude ($40^\circ$ to $69^\circ$), imposing constraints on coronal models for M dwarfs. 

We have noted that both ExPRES simulations (constrained by the ECM energy source, characteristic electron energy in the case of a loss-cone, and L-shell) and drift-rate estimates (determined by the electron energy, L-shell, and electron pitch angle at the equator) gave consistent results, pointing at radiating electrons with energy E=20-30 keV moving along field lines with L$\sim 2-10$. The above plasma density estimates rather favour the range L$\sim 5-10$. We don't claim a high accuracy on either L or E, but the results are indicative of mid- to high-latitude emissions by moderately energetic electrons. Positive drifts on December $2^{nd}$ rather favour shell-driven ECM with downgoing electrons (in that case along L$\sim$2 field lines, while loss-cone is compatible with the negative drift-rates observed on December $3^{rd}$. On this latter day, drift-rate calculations suggest that the bursts are produced either by a mechanism different from ECM -- the overall drifts being either coincidental, or due to an overall source motion at a speed similar to that of electrons in adiabatic motion along dipolar field lines \citep{Willes2002} --, or along magnetic field lines not described by a large-scale dipolar field (e.g. smaller scale magnetic loops).

These results, based on two 3-hour observations only, are of course preliminary, but they are a first quantitative analysis\pz{, a proof of concept that shows that detailed characterization of ECM emission regions, electron energies, and coronal plasma density becomes possible using fine structures observed in stellar radio bursts}. 

The small t-f extent of the observed overall t-f patterns being much less extended than our simulations of Figure \ref{envelopes} suggests that restricted longitude ranges, likely different, are emitting on the two days. More observations, especially clustered in time, will obviously bring better constraints. At a finer timescale, resolved stellar radio bursts also carry information about their magneto-plasma of origin and can be effective diagnostic tools for the emission mechanism and electron acceleration process. 

It must be noted that ExPRES simulations and drift-rate calculations benefit from the knowledge acquired on ECM operation in solar system planetary magnetospheres \citep{Zarka1998,Treumann2006}. For example, the quasi-periodicity at $\sim$5 Hz of burst occurrence on December $2^{nd}$, 2021 suggests intermittent electron acceleration. At Jupiter, similar bursts (Figure \ref{comparisons}) were interpreted as due to electrons accelerated by Alfvén waves of frequency 5-20 Hz, excited along Jupiter's magnetic field lines by its moon Io, and amplified at the feet of these field lines in the so-called ionospheric Alfvén resonator \citep{Hess2007,Mauduit2023}. Alfvén waves have also been invoked in the case of Solar spikes \citep{Wu2007}. Constraints posed by radio burst observations may thus eventually constrain the Alfvén speed ($V_A = c (f_{ce}/f_{pe}) (m_e/m_p)^{1/2}$, with $c$ the speed of light, and $m_e$ and $m_p$ the masses of the electron and proton, respectively) at the base of the corona.

Based on the above conclusions, we may speculate a little further about the large-scale physical driver of AD Leo's radio bursts, in particular do they result from the flare paradigm \citep[see e.g.,][]{Zic2020} or from the planetary magnetosphere paradigm \citep[see e.g.,][]{Hallinan2015}.
The flare paradigm leaves room for ECM emission as long as adequate $f_{pe}/f_{ce}$ ratios exist in the source region \citep{Morosan2015,Morosan2016,Yu2024}. The similarity of the spikes from December $3^{rd}$ with Solar spikes (Figure \ref{comparisons}) rather supports this paradigm, although no evident correlation was noted in paper \#1 between optical and radio flares (however optical flares might be too small to be detected by the optical telescopes used in paper \#1, or they could happen before the radio flares and inject accelerated electrons into the large-scale magnetic field, whose radio signature would occur after some accumulation \citep[e.g.,][]{Yu2024}).
In the planetary magnetosphere paradigm, the main question is the source of electron acceleration (or of the Alfvén waves that cause it). Usual suspects are corotation breakdown and star-planet interaction with a planetary companion. We discussed in Sections \ref{sec:AD_Leo_MFL_chara} and \ref{sec:ExPRES_analysis} the unlikeliness of an adequate planetary companion for AD Leo, and that is the reason why we did not explore it via ExPRES simulations beyond the synchronous orbit (this would have involved too many free parameters). Further observations may lead us to reconsider this possibility. We are thus left with corotation breakdown \citep{Nichols2012,Turnpenney2017}, and with the fact that we located the radio sources along field lines of relatively low L, i.e. likely in the sub-Alfvénic region of the stellar corona. Considering the more massive wind of red dwarfs and the $\sim$13 times faster rotation of the star compared to the Sun's, we propose that electron acceleration driving ECM on sets of field lines in restricted longitude sectors may be caused by corotation breakdown applied to plasma blobs in the inhomogeneous stellar wind, occurring at a few $R_*$ from the star, similar to what happens at Jupiter in the external regions of the Io plasma torus \citep{Yang1994,Louarn1998}.

\section{Perspectives}
\label{sec:perspectives}

With the start of operations of the large sensitive low-frequency array NenuFAR in France, in the range 10-85 MHz \citep{Zarka2020}, low-frequency observations of AD Leo's radio bursts might bring complementary constraints to those obtained at higher frequencies with FAST. 

\begin{figure}[!ht]
	\begin{center}
	\centerline{\includegraphics[width=1.0\linewidth]{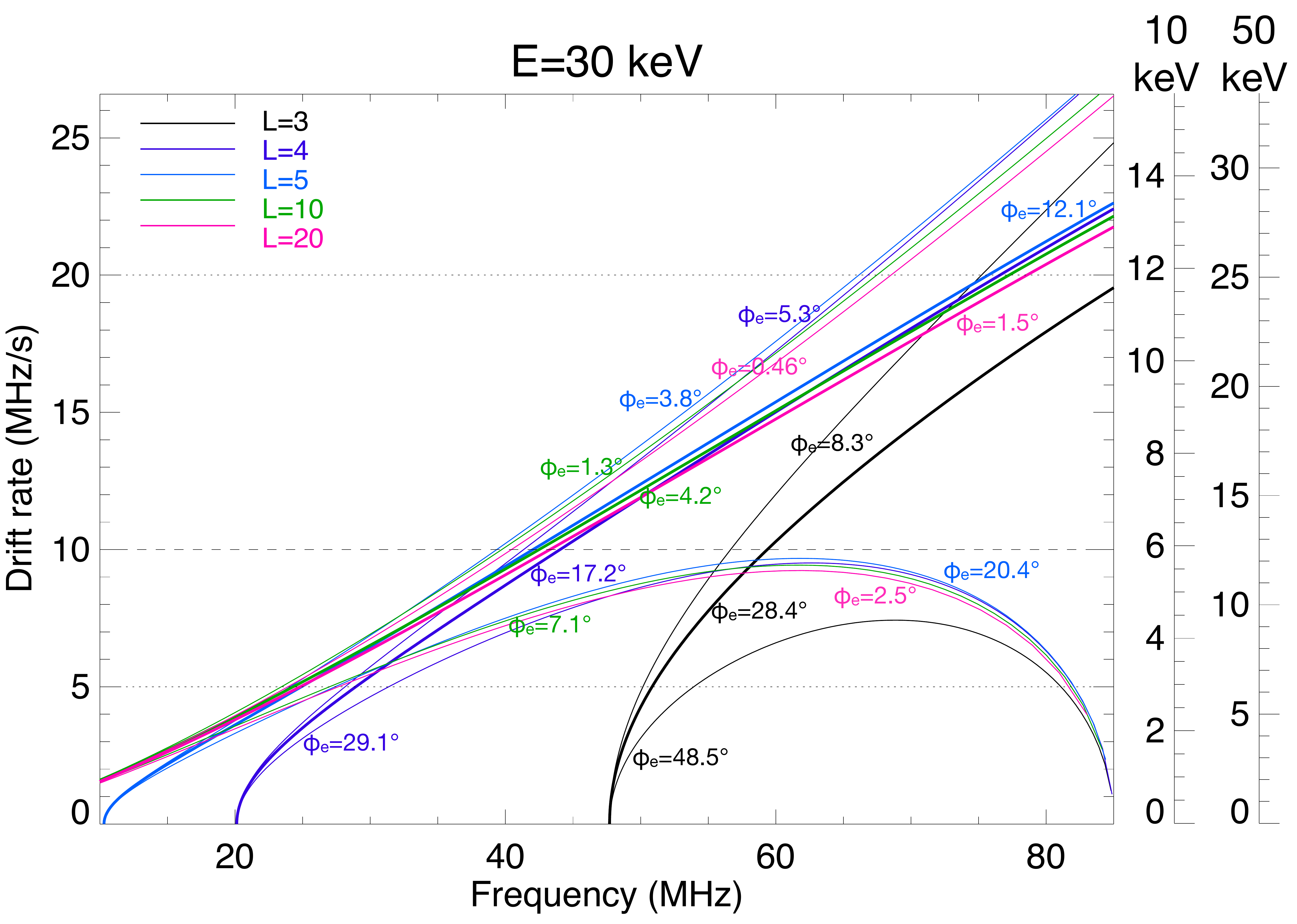}}
\caption{Drift-rates similar to Figure \ref{drifts_calca} but calculated in NenuFAR's frequency range, for an electron energy of 30 keV. The curves are globally shifted vertically by a change of electron energy, hence the different vertical scales on the right side corresponding to different electron energies of 10 and 50 keV.}
\label{drifts_calc_n}
\end{center}
\end{figure}

In support of such observations, that have actually started in 2023, we have extrapolated Figure \ref{drifts_calca} in the spectral range of NenuFAR in order to estimate the magnitude of the drift-rates that should be searched for. Those are displayed in Figure \ref{drifts_calc_n}. Surprisingly enough, we find drift-rates of order of 10 MHz/s, very similar to those of Jupiter S-bursts in the same spectral range. We note that cyclotron frequencies matching the lowest frequencies of the NenuFAR range cannot be reached in AD Leo's environment on field lines with L$<$5 ($f_{ce} \ge 20$ MHz along L=4, $f_{ce} \ge 48$ MHz along L=3, and $f_{ce} \ge 85$ MHz along L=2).

\begin{figure}[!ht]
	\begin{center}
	\centerline{\includegraphics[width=1.0\linewidth]{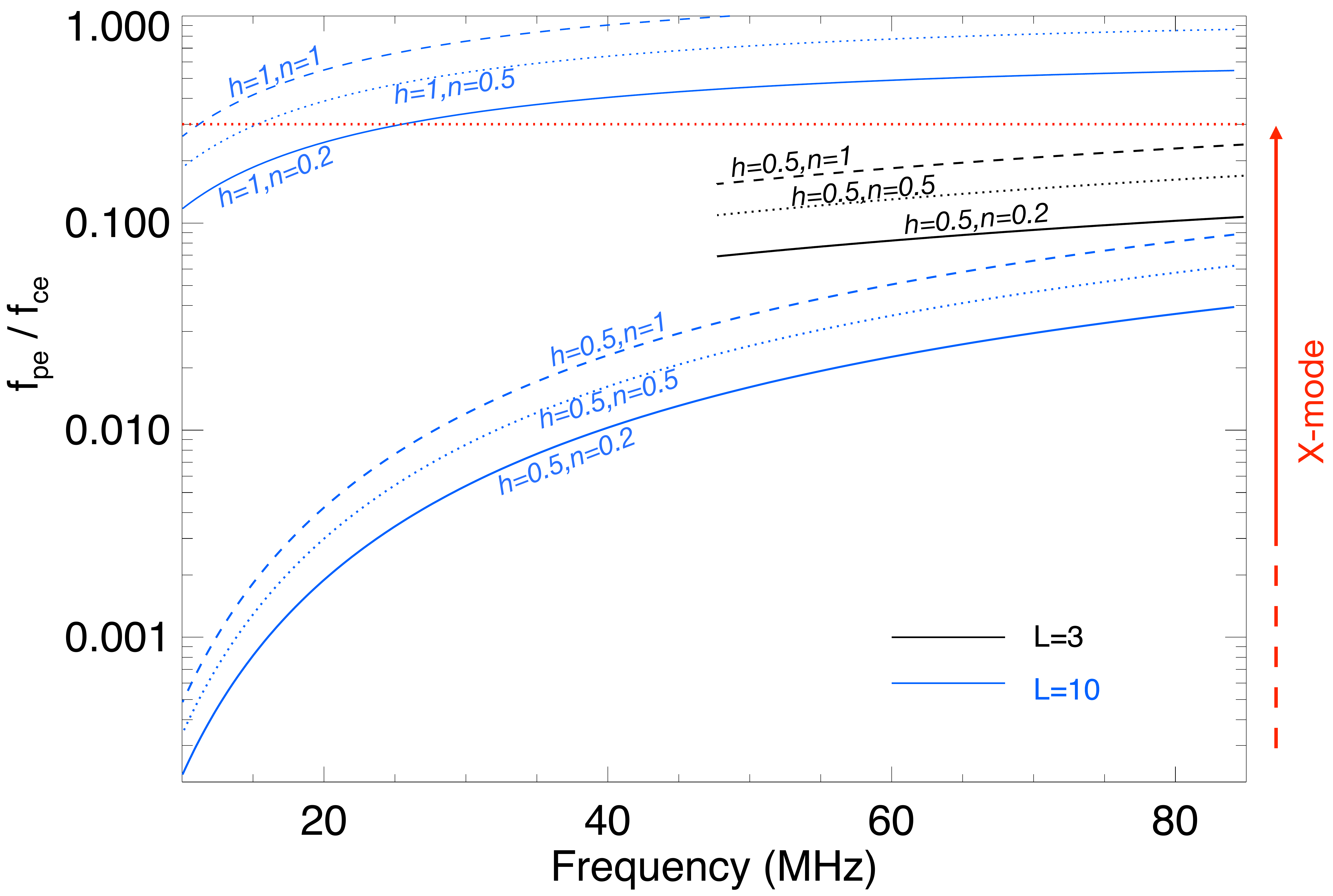}}
\caption{$f_{pe}/f_{ce}$ values computed as in Figure \ref{fpe_fce} but in the frequency range of NenuFAR.}
\label{fpe_fce_n}
\end{center}
\end{figure}

Similarly, we have extrapolated Figure \ref{fpe_fce} in the spectral range of NenuFAR and obtained Figure \ref{fpe_fce_n}. If the hydrostatic description of AD Leo's atmosphere holds up to a few $R_*$ distance, low values of $f_{pe}/f_{ce}$ allowing for fundamental R-X mode ECM emission (that can be identified through its circular polarization sense) should exist in the low-frequency source region for a relative coronal scale height twice lower than in \citet{Villadsen2019} ($h \leq 0.5$). 

Observations of emission occurrence and drift-rates at low radio frequencies would thus be very powerful for better locating the sources and estimating electrons' energies. Coordinated high- and low-frequency observations are also potentially very informative, as Figure \ref{expres} shows that due to the ECM beaming combined with the stellar rotation, emission from the same field line at different frequencies is expected to be received on Earth with a delay that may reach several hours. The measured delay will be a very strong constraint on the fit to ExPRES simulations. Of course, contemporaneous ZDI maps will be of utmost importance to apply ExPRES to reliable magnetic field topologies.

\begin{acknowledgements}
P.Z. and C.L. acknowledges funding from the ERC under the European Union's Horizon 2020 research and innovation program (grant agreement N$^\circ$ 101020459—Exoradio).
J.Z. and H.T. are supported by NSFC grant 12250006. This work made use of the data from FAST (Five-hundred-meter Aperture Spherical radio Telescope).
FAST is a Chinese national mega-science facility, operated by National Astronomical Observatories, Chinese Academy of Sciences.
\end{acknowledgements}

%
%


\bibliography{adleo}


\begin{appendix}
\setcounter{figure}{0}
\renewcommand\thefigure{\thesection.\arabic{figure}}
\onecolumn
\section{Drift-rates of sub-bursts and sub-burst trains on December 3$^{rd}$, 2021}
\label{appendix:overall_drift_rate}

On December $2^{nd}$, drift-rates of linear t-f structures were easily measurable. Figure 3a of paper \# 1 and Figure \ref{drifts_obs}c of the present paper showed the clear dependence of $df/dt$ on the frequency, varying from +550 MHz/s at 1000 MHz to +970 MHz/s at 1470 MHz. 

\begin{figure}[h]
	\begin{center}
	\centerline{\includegraphics[width=0.85\textwidth]{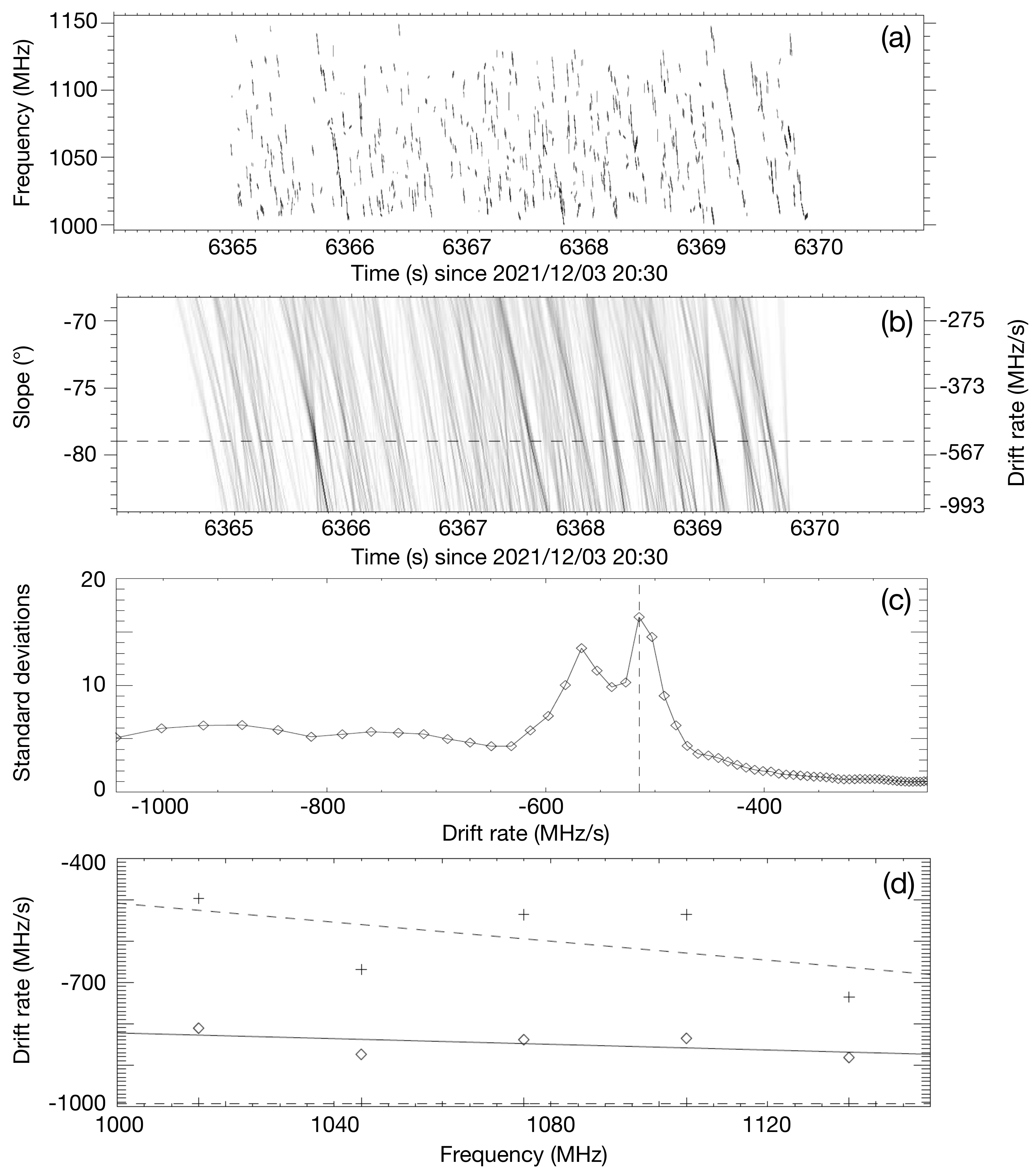}}
\caption{Example of determination of the overall drift-rate of sub-burst trains on December 3$^{rd}$, 2021. (a) Five seconds of catalogued sub-bursts reproduced from Figure 8c of paper \#1. (b) Result of parametric linear dedispersion and spectral integration of (a), as a function of slope (left scale) or drift-rate (right scale). The dashed horizontal line about -500 MHz/s indicates the drift-rate for which most of the burst clusters of (a) align. (c) Standard deviation of each line of (b) (actually exponential of values of (b) are used to enhance the contrast). The dashed vertical line indicates the peak value, obtained at -514 MHz/s. (d) Distribution of overall drift-rates computed in 20 s $\times$ 30 MHz intervals between 1000 and 1150 MHz. For each of the 5 frequency intervals, the diamond indicates the average overall drift-rate and the `+' the 10 \% and 90\% quantiles of their distribution. The solid line is the linear fit of the average values, and the dashed lines those of the quantiles. The lower values about -1000 MHz/s are `saturated' by the time resolution of the original dynamic spectrum (a), that does not allow to distinguish clearly between drift-rates steeper than -1000 MHz/s, thus we limited our parametric dedispersion to that value.}
\label{drifts_3dec}
\end{center}
\end{figure}

On December $3^{rd}$, the morphology of the bursts, very different from the previous day, was blob- or spot-like as shown in Figure \ref{drifts_obs}b, reminiscent of some solar radio spikes \citep{Benz1986,Wu2007,Chernov2008} (see also Figure \ref{comparisons}c,d,e). 
About 50000 individual sub-bursts were catalogued, with drift-rates from -5000 to + 3300 MHz/s (dots and `+' in Figure \ref{drifts_obs}c). The linear fit of $df/dt(f)$ leads to the short-dashed orange line on Figures \ref{drifts_obs}, \ref{drifts_calca} and \ref{drifts_calcb}, that is not compatible with electron's adiabatic motion in a large-scale dipolar magnetic field (cf. Figures \ref{drifts_calca} and \ref{drifts_calcb}).
But in paper \#1, sub-burst alignments or clusters were clearly identified, as illustrated in Figures \ref{drifts_obs}b and \ref{drifts_3dec}a (the latter reproduces Figure 8c of paper \#1).

In order to quantify the overall drifts of these clusters, we performed a re-analysis of the sub-burst catalogue. This catalogue contains for each sub-burst its t-f position and linear shape as well as its intensity and polarization. Figure \ref{drifts_3dec}a displays the corresponding sub-bursts detected in the same interval as Figure 8c of paper \#1. To obtain an estimate of the overall drift of sub-burst trains, we ``dedispersed'' the dynamic spectrum as is done for pulsar pulses \citep[e.g.,][]{Zakharenko2013}, but correcting for a linear drift instead of one in $1/f^2$ drift for pulsars (i.e. for each trial drift-rate $df/dt$, we shifted each time series at frequency $f$ relative to a reference frequency $f_0$ by $\delta t = df/ft \times (f_0 - f)$), and we integrated the dedispersed signal to obtain a time series for each trial drift-rate. The result of this operation is displayed on Figure \ref{drifts_3dec}b as a function of time and slope or drift-rate. In the 5 s example displayed, the overall drift-rate clearly peaks around -500 MHz/s (horizontal dashed line), where most caustics appear in the $t - df/dt$ plane. 
To determine automatically the best drift-rate, we computed the standard deviation of the time series obtained for each trial drift-rate, displayed in Figure \ref{drifts_3dec}c (where we have taken the exponential of the values in Figure \ref{drifts_3dec}b in order to enhance the contrast of the result). It peaks for the drift-rate that aligns best the sub-burst clusters, here equal to -514 MHz/s (vertical dashed line). 
We have performed this analysis in frequency intervals of 30 MHz and time slices of 20 s over the entire observation of December $3^{rd}$. In each frequency interval we computed the average overall drift-rate and the 10\% and 90\% quantiles of their distribution, plotted in Figure \ref{drifts_3dec}d. The variation of the 10\% and 90\% quantiles (dashed) define the lower grey-shaded area in Figure \ref{drifts_obs}c, and variation of the the average overall drift-rate with frequency (solid line) is the long-dashed orange line in Figures \ref{drifts_obs}, \ref{drifts_calca} and \ref{drifts_calcb}.

\setcounter{figure}{0}
\renewcommand\thefigure{\thesection.\arabic{figure}}
\section{General derivation of the ECM radio beaming angle $\theta$ for the R-X and L-O modes}
\label{appendix:dispersion_equation_solution}

We detail here how to derive
the dispersion equation in a cold magnetized plasma, and its solution as a function of $\theta$, the angle between the local magnetic field $B$ and the wave vector $k$.


\subsection{Dispersion equation}
To obtain the dispersion equation in a cold magnetised plasma, we use the following Maxwell equations \citep{Maxwell1865}:\\
the Maxwell-Amperes relationship:
\begin{equation}
\vec{\nabla}\times{B} = \mu_0\vec{J}+\epsilon_0\mu_0 \frac{\partial{\vec{E}}}{\partial{t}}
\end{equation}
the Maxwell-Faraday relationship:
\begin{equation}
\vec{\nabla}\times\vec{E}=-\frac{\partial{\vec{E}}}{\partial{t}}
\end{equation}
the current density formula:
\begin{equation}
\vec{J}=\tensorsym{\sigma} \vec{E}
\end{equation}
and that of the dielectric tensor:
\begin{equation}
\tensorsym{K}=\tensorsym{1}-\frac{\tensorsym{\sigma}}{i\omega\epsilon_0}
\end{equation}

We will also use the fact that solutions are in the form of plane waves, and that all the quantities $f(\vec{r},t)$ are proportional to $e^{i(\vec{k}.\vec{r}-\omega t)}$.

That gives us:
\begin{equation}
\frac{c^2}{\omega^2} \vec{k}\times\vec{k}\times\vec{E}+\tensorsym{K}\vec{E} = 0
\end{equation}

The refractive index $N$ of a medium is defined as the ratio of the speed of light in a vacuum $c$ to the phase velocity $v_\phi$ of a wave propagating in the medium. In the case of a plane wave, the phase velocity corresponds to the propagation speed of the wavefront along the wave vector $\vec{k}$, i.e. $v_\phi = \frac{w}{k}$. Therefore, $\vec{N}=\frac{c\vec{k}}{\omega}$.

The magnetic field $\vec{B}$ (of unit vector $\vec{b}$) is directed along $\vec{e_z}$. The wave vector $\vec{k}$ is contained in the plane $(\vec{e_x}, \vec{e_z})$. Theses two vectors form an angle $\theta$. In this reference system, we can define $\vec{N}=(N sin{\theta},~0,~N cos{\theta})$.

We therefore obtain:
\begin{equation}
\frac{c^2}{\omega^2} \vec{k}\times\vec{k}\times\vec{E} =
\left (
\begin{array}{ccc}
-N^2 cos^2{\theta} & 0 & N^2 cos{\theta}sin{\theta} \\
0 & -N^2 & 0 \\
N^2 cos{\theta} sin {\theta} & 0 & -N^2 sin^2{\theta}
\end{array}
\right )
\left (
\begin{array}{c}
E_{x}\\
E_{y} \\
E_{z}
\end{array}
\right )
\end{equation}

For deriving  $\tensorsym{K}\vec{E}$, we start from the equation of motion:

\begin{equation}
\sum \vec{F}=  m \frac{\partial \vec{v}}{\partial t} =e[\vec{E}+\vec{v}\times\vec{B}]
\end{equation}
giving us:
\begin{equation}
m_s \frac{\partial \vec{v_s}}{\partial t} =e_s[\vec{E_1}+\vec{v_{s1}} \times \vec{B_0}]
\end{equation}
with $s$ standing for ``species'' (ion or electron),
$\vec{E}= \vec{E_0}+ \vec{E_1}= \vec{E_1}$,
$\vec{v_s}= \vec{v_0}+ \vec{v_{s1}}= \vec{v_{s1}}$ (cold plasma),
$ \vec{B}= \vec{B_0}+ \vec{B_1}$,
and $\vec{v_{s1}} \times \vec{B_1} = O^2(\epsilon)$ (second order, very low in front of first order).

As we are assuming the solutions are in the form of plane wave, by projecting along x, y and z, we obtain:

\begin{equation}
\left (
\begin{array}{ccc}
-i \omega&-\omega_{cs}&0 \\
-\omega_{cs}&-i \omega&0 \\
0&0&-i \omega
\end{array}
\right )
\left (
\begin{array}{c}
v_{sx}\\
v_{sy} \\
v_{sz}
\end{array}
\right )
=\frac{e_s}{m_s} 
\left (
\begin{array}{c}
E_{x}\\
E_{y} \\
E_{z}
\end{array}
\right )
\end{equation}
With $\omega_{cs}=\frac{eB_0}{m_s}$ the cyclotron pulsation for each species. Hence

\begin{equation}
\left (
\begin{array}{c}
v_{sx}\\
v_{sy} \\
v_{sz}
\end{array}
\right )
=\frac{e_s}{m_s} 
\left (
\begin{array}{ccc}
\frac{-i \omega}{\omega_{cs}^2-\omega^2}&\frac{\omega_{cs}}{\omega_{cs}^2-\omega^2}&0 \\
\frac{-\omega_{cs}}{\omega_{cs}^2-\omega^2}&\frac{-i \omega}{\omega_{cs}^2-\omega^2}&0 \\
0&0&\frac{i \omega}{\omega}
\end{array}
\right )
\left (
\begin{array}{c}
E_{x}\\
E_{y} \\
E_{z}
\end{array}
\right )
\end{equation}

The current density writes:

\begin{equation}
\vec{J} = \sum_{s} n_{s0} e_s \vec{v_s} = \tensorsym{\sigma} \vec{E}
\end{equation}
from which we obtain the conductivity tensor $\tensorsym{\sigma}$ :
\begin{equation}
\tensorsym{\sigma}=\sum_{s} \frac{e^2n_{s0}}{m_s}
\left (
\begin{array}{ccc}
\frac{-i \omega}{\omega_{cs}^2-\omega^2}&\frac{\omega_{cs}}{\omega_{cs}^2-\omega^2}&0 \\
\frac{-\omega_{cs}}{\omega_{cs}^2-\omega^2}&\frac{-i \omega}{\omega_{cs}^2-\omega^2}&0 \\
0&0&\frac{i \omega}{\omega}
\end{array}
\right )
\end{equation}

Then we can express the dielectric tensor:

\begin{equation}
\tensorsym{K}=\tensorsym{1}-\frac{\tensorsym{\sigma}}{i\omega\epsilon_0}
\end{equation}
which is therefore written:
\begin{equation}
\tensorsym{K}=
\left (
\begin{array}{ccc}
S&-iD&0\\
iD&S&0\\
0&0&P
\end{array}
\right )
\end{equation}
where S, P and D correspond to the Stix notation \citep{Stix1962}:

\begin{equation} \label{coefficientsSPD}
\begin{split}
S&=1-\sum_{s}\frac{\omega_{ps}^2}{\omega^2-\omega_{cs}^2} \\
P&=1-\sum_{s}\frac{\omega_{ps}^2}{\omega^2} \\
D&=\sum_{s}\frac{\omega_{ps}^2\omega_{cs}}{\omega(\omega^2-\omega_{cs}^2)}
\end{split}
\end{equation}
with $\omega_{ps}^2=\frac{n_s e^2}{\epsilon_0 m_s}$.\newline 

Finally:

\begin{alignat}{2}
\begin{split}
\frac{c^2}{\omega^2} \vec{k}\times\vec{k}\times\vec{E}+\tensorsym{K}\vec{E} & = 0 \\
\Longleftrightarrow
\left (
\begin{array}{ccc}
S-N^2cos^2{\theta}&-iD&N^2 cos{\theta} sin {\theta}\\
iD&S-N^2&0\\
N^2 cos{\theta} sin {\theta}&0&P-N^2 sin^2 {\theta}
\end{array}
\right )
\left (
\begin{array}{ccc}
E_{x}\\
E_{y} \\
E_{z}
\end{array}
\right )
=0
\end{split}
\end{alignat}

The determinant of this matrix gives an N-squared equation of the form:

\begin{equation}
AN^4+BN^2+C = 0
\label{relation_dispersion}
\end{equation}
with
\begin{equation} \label{coefficientsABC}
\begin{split}
A&=P cos^2 \theta + S sin^2 \theta \\
B&=sin^2 \theta (D^2-S^2) - SP (1+ cos^2 \theta)\\
C &= P(S^2-D^2)
\end{split}
\end{equation}
In oblique propagation (i.e. $\theta \neq 0 \degree$ or $90 \degree$) , the solutions are given by:

\begin{equation}
\label{eq:solutions_dispersion_equation_N}
N^2 = \frac{-B \pm \sqrt{B^2 - 4 AC}}{2A}
\end{equation}

The solutions of this dispersion equation give a relation between the refractive index $N$, the wave gyrofrequency $\omega$ and the angle $\theta$ between the wave vector and the ambient magnetic field. 

The $\pm$ sign indicates the existence of two branches to the solutions. To understand the meaning of this $\pm$ for the wave, we derive in the next Section the Altar-Appleton-Hartree expression of the refractive index.

\subsection{Altar-Appleton-Hartree expression}
To find the solution to this determinant directly in Altar-Appleton-Hartree form, it is easier to put $N^2=1+\xi$ and solve
\begin{equation}
A + \frac{2A-B}{\xi} +  \frac{A-B+C}{\xi^2}=0
\end{equation}

The solutions of this equation are given by:
\begin{equation}
N^2 = 1+ \frac{2(A-B+C)}{-2A+B \pm \sqrt{B^2-4AC}}
\end{equation}

In the following, we will use the approximation that at high frequencies, the terms relating to the ions in the expression of $S$, $P$ and $D$ are negligible, because $m_i >> m_e$ and therefore $\omega_{pi}<<\omega_{pe}$, as well as $\omega_{ci}<<\omega_{ce}$

To finally arrive at the Altar-Appleton-Hartree equation, each term in the above equation must be rewritten, factoring by $\left ( \frac{\omega_{pe}^2}{\omega^2-\omega_{ce}^2}\right )^2 \frac{1-\frac{\omega_{ce}^2}{\omega^2}}{\frac{\omega_{pe}^2}{\omega^2}}$

In the end, this gives us:
\begin{equation}
N^2=1-\frac{2*\frac{\omega_{pe}^2}{\omega^2}*(1-\frac{\omega_{pe}^2}{\omega^2})}{2*(1-\frac{\omega_{pe}^2}{\omega^2})-\frac{\omega_{ce}^2}{\omega^2}sin^2{\theta} \pm \sqrt{(\frac{\omega_{ce}^2}{\omega^2}*sin^2{\theta})^2 + 4*\frac{\omega_{ce}^2}{\omega^2}*cos^2{\theta}*(1-\frac{\omega_{pe}^2}{\omega^2})^2}}
\label{Ngeneraleannexe}
\end{equation}

\begin{figure}[!ht]
	\begin{center}
	\centerline{\includegraphics[width=0.9\linewidth]{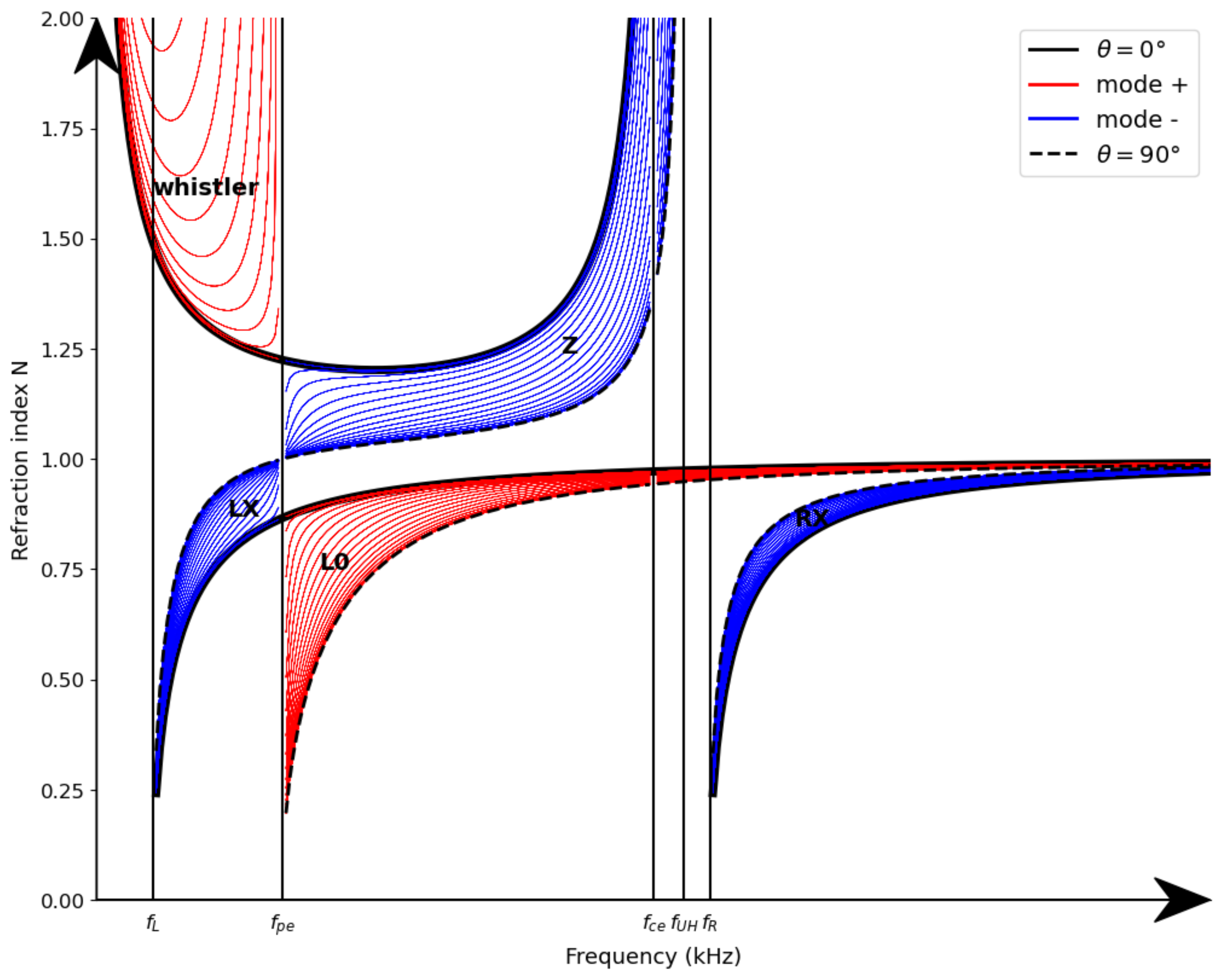}}
\caption{Refractive Index N as a function of the frequency f ($\omega = 2 \pi f$) for low-density plasma (i.e., $f_{ce}>f_{pe}$, requires for the ECM to occur). The red areas corresponds to the $+$ branch of the Appleton-Hartree expression, while the blue area correspond to the $-$ branch of the Appleton-Hartree expression. The oblique propagation dispersion curves for different $\theta$ are shown as thin coloured lines ($\delta \theta = 5\degree$), while dispersion curves for parallel propagation ($\theta = 0\degree$) are shown as thick black lines, and those for perpendicular propagation ($\theta = 90 \degree$) as thick black dashed lines.
The L-mode cutoff $f_L$, plasma cutoff $f_{pe}$, cyclotron resonance $f_{ce}$, upper hybrid resonance $f_{UH}$ and R-mode cutoff $f_R$ frequencies are indicated on the x-axis. In this example, $f_{ce}/f_{pe} = 0.3$. In reality, the $f_{ce}/f_{pe}$ is lower than that, but has been enlarged here for the sake of readability.}
\label{Brillouin_diagram}
\end{center}
\end{figure}

As mentioned earlier, the $\pm$ sign indicates the existence of two branches to the solutions. These branches (represented Figure \ref{Brillouin_diagram}) are framed by cutoff and resonance frequencies and named as follow:
\begin{itemize}
    \item[$+$]: \textbf{whistler} mode ($\omega < \omega_p$);\\
                ~~Left-Ordinary (\textbf{LO}) mode ($\omega > \omega_p$).
    \item[$-$]: Left-eXtraordinary (\textbf{LX}) mode ($\omega_L < \omega < \omega_p$);\\
                ~~Right-eXtraordinary(RX)-\textbf{Z} mode ($\omega_p < \omega < \omega_{UH}$);\\
                ~~\textbf{RX} mode ($\omega > \omega_R$).\\        
\end{itemize}

Note that below $\omega_{ce}$, in oblique and parallel propagation (so for $\theta < 90\degree$), the naming of the branches of the Appleton-Hartree equation from the right and left modes depends on the value of the frequency, as there is a discontinuity in $\omega = \omega_{pe}$, which implies a change of sign in the equation. Therefore:
\begin{itemize}
    \item for $\omega > \omega_{ce}$~$\rightarrow$~$N_R = N_+$ and $N_L = N_-$
    \item for $\omega_{ce} > \omega > \omega_{pe}$~$\rightarrow$~$N_R = N_-$ and $N_L = N_+$
    \item for $\omega_{pe} > \omega$~$\rightarrow$~$N_R = N_+$ and $N_L = N_-$
\end{itemize}
meaning that in-between $\omega_{pe}$ and $\omega_{ce}$ there is an inversion of the branches to the solutions for the right and left modes (this is taken into account in Figure \ref{Brillouin_diagram}).

In Figure \ref{Brillouin_diagram} we see that only L-O and R-X modes tend to $N=1$ and therefore tend towards a high-frequency light wave. These modes are therefore capable of propagating in the vacuum outside of the plasma. It is these two modes that we simulate in ExPRES.

\subsection{Solution of the Dispersion equation function of $\theta$ in the oblique case (i.e. loss-cone distribution function)}

We now want the solution of the dispersion equation (see Equation \ref{eq:solutions_dispersion_equation_N}), but only function of $\theta$, the angle between the ambient magnetic field $B$ and the wave vector $k$, as this is what we compute numerically, and in a self-consistent way, in the ExPRES code. 

\subsubsection{Equation of the Electron Cyclotron Maser Instability in the oblique case (i.e. loss-cone distribution function)}
In the Electron Cyclotron Maser Instability case, the oblique propagation is obtained when the instability is driven by a loss-cone electron distribution function. The wave-particle resonance equation for the Electron Cyclotron Maser Instability writes:
\begin{equation}
\omega - k_{||} v_{r ||} = \omega_{\text{ce}} \Gamma_r^{-1},
\label{resonance}
\end{equation}
where $\omega_{\text{ce}}$ is the electron cyclotron angular frequency, $||$ refers to the parallel component of the wave vector $k$ and of the velocity of the resonant ($r$) electron $v_{r}$, and $\Gamma_r$ is the Lorentz factor associated with the gyration motion of the electrons:
\begin{equation}
\Gamma_r^{-1}=\sqrt{1-\frac{v_r^2}{c^2}}.
\end{equation}

We remind that the aperture of the conical emission sheet $\theta$ is defined by the angle between the magnetic field vector and the wave vector. The magnetic field $\vec{B}$ (of unit vector $\vec{b}$) is directed along $\vec{e_z}$. The wave vector $\vec{k}$ is contained in the plane $(\vec{e_x}$;$\vec{e_z})$.  Assuming that the emission pattern has a cylindrical symmetry of revolution around the magnetic field line, the angle $\theta$ can be defined as :

\begin{equation}
k cos \theta = \vec{k} \vec{b}  = k_{||}.
\label{def_eq_theta}
\end{equation}

In the weakly relativistic case, the resonance equation is a circle in velocity space:
\begin{equation}
v_\perp ^2+(v_{||}-v_0)^2 	\approx c^2 \left(\frac{k_{||}^2 c^2}{\omega_{ce}^2} + 2 \left(1- \frac{\omega}{\omega_{ce}} \right)\right).
\label{resonance_cercle}
\end{equation}

\begin{align}
& \text{with center: } \left( v_{\perp_{0}} = 0 \text{ ; } v_{||_{0}} = v_0 = \frac{k_{||} c^2}{\omega_{ce}} \right) \label{v0_1},\\
& \text{and radius: } R=c \left(\frac{k_{||}^2 c^2}{\omega_{ce}^2} + 2 \left(1- \frac{\omega}{\omega_{ce}} \right)\right)^{1/2}.
\label{R}
\end{align}

From Equations \ref{def_eq_theta} and \ref{v0_1}, and by inserting the refractive index of the medium which is written:

\begin{equation}
    N=ck/\omega
\end{equation}

$v_0$ can be rewritten as follows:

\begin{equation}
v_{{||}_0} = v_0 = c N \frac{\omega}{\omega_{ce}} cos \theta.
\label{v0_2}
\end{equation}

\begin{figure}[H]
	\center
    \includegraphics[width=0.9\textwidth]{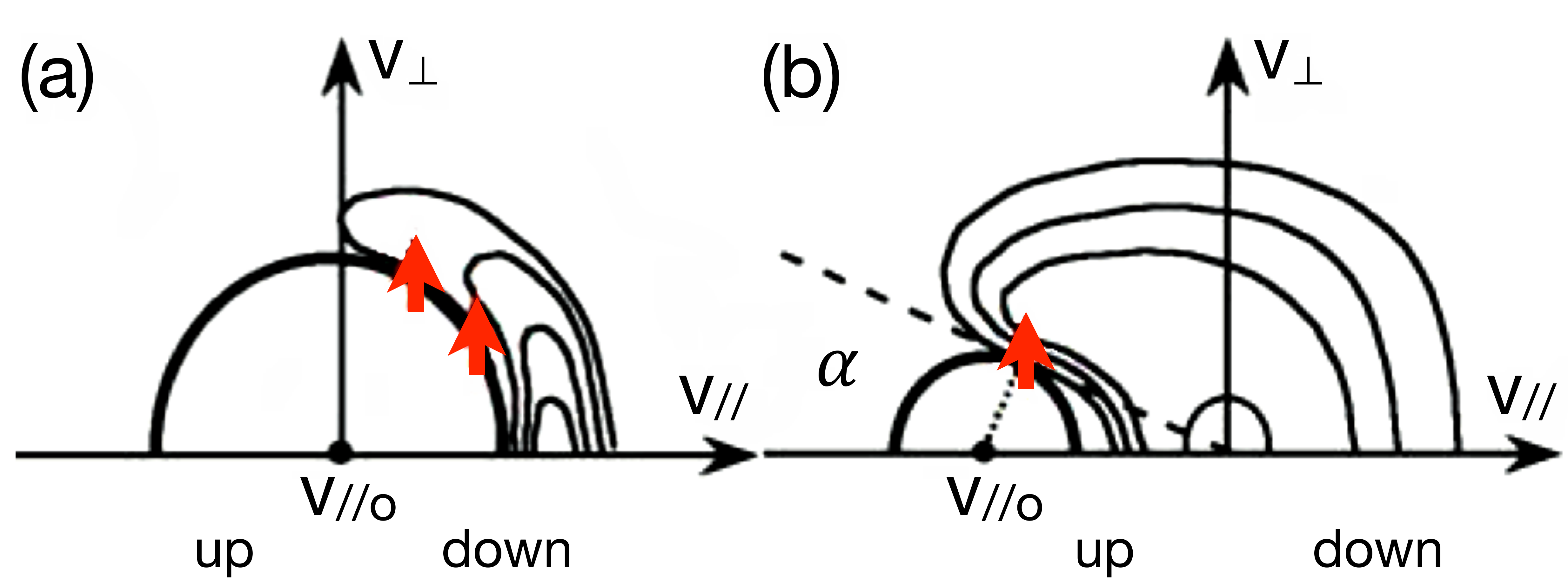}
\caption{(a) Shell-type electron distribution. The $\nabla v_\perp$ positive gradient of the distribution function is located along the outer edge of the distribution. The ECM resonance circle leading to amplification is tangent to this and is centered at $v_{||_0} = 0$; (b) loss-cone electron distribution. The loss-cone is a region depleted in electrons,  lost by collision with the atmosphere. This region is located below the dashed line, defined by the pitch angle $\alpha$. The largest positive gradient $\nabla v_\perp$ of the distribution function (red arrow) is at the edge of the loss-cone,  at resonance velocity $v$. The ECM resonance circle leading to maximum amplification is tangent to this gradient, and is centered on $v_{||_0} = v_{r}/cos \alpha$ -- Figure adapted from \citet{Hess2008}.}
\label{horseshoe}
\end{figure}

For loss-cone--type electron distributions,  the resonance circle (Figure \ref{horseshoe}b) is centered on a non-zero velocity $v_0$ (see Equations \ref{v0_1} and \ref{v0_2}), which produces an oblique emission ($k_{||} \neq 0$) with respect to the magnetic field (see Equation \ref{resonance}). The position of the resonance circle is related to the angle of the loss-cone $\alpha$, as shown in Figure \ref{horseshoe}b. This pitch angle defines the position of the mirror point (the point where $v_{||}=0$):

\begin{equation} \label{angle_attaque}
cos \alpha = \frac{v_{||}}{v} = \left( 1-\frac{\omega_{ce}}{\omega_{ce_{max}}}\right)^{-1/2} \text{ and } sin \alpha = \frac{v_\perp}{v},
\end{equation}
with $\omega_{ce_{max}}$ the cyclotron pulsation at the mirror point, i.e. here at the altitude of the peak of the UV aurora. Jupiter's UV aurorae are indeed the result of the collision between the energetic electrons (associated with the radio emissions considered here) and the neutral atmosphere. Their altitude is therefore a good approximation of the limit mirror point, beyond which the electrons are lost in the atmosphere, producing the cone of loss visible in Figure \ref{horseshoe}b.

\subsubsection{Determination of the solution of the Dispersion equation function of $\theta$}
As we want the solution of the dispersion equation (see Equation \ref{eq:solutions_dispersion_equation_N}), but only function of $\theta$, we therefore go back to Equation \ref{relation_dispersion}.
So we first express the refractive index $N$ specific to the loss-cone electron distribution function. 

For that, we use several equations: Equation \ref{v0_2} which describe the center $v_0$ of the resonant circle as a function of $N$ and $\theta$; the expression of the resonant electron energy $v_r$ and $v_0$ as a function of the pitch angle $\alpha$ (that describes the position of the mirror point where $v_{||} = 0$):
\begin{equation}
    v_0 cos \alpha = v_r
\end{equation} 
(see Figure \ref{horseshoe} a for resonant electrons)
and finally, using the fact that $v_r << c$ (which is true for electron energy of a few keV to $\sim 30$~keV) , we can express the value of $\omega$ for oblique emissions only as a function of $v_r$:

\begin{align*}
    \omega_{lc} &= w_{ce} \Gamma _r^{-1} + \frac{v_0 \omega_{ce}}{c^2} ~ v_{r||}&\\
    &= w_{ce} \Gamma _r^{-1} + \frac{v_0 ~\omega_{ce}}{c^2} ~ v_r ~ cos \alpha &\text{ ~~~~~~~~ cf Equation \ref{angle_attaque}} \\
&= w_{ce} \left( \Gamma _r^{-1} + \frac{v_r^2}{c^2} \right) &\text{ ~~~~~~~~ using $v_0 ~ cos \alpha = v_r$, cf. Figure \ref{horseshoe}} \\
& = w_{ce} \left( (1 - \frac{v_r^2}{c^2})^\frac{1}{2} + \frac{v_r^2}{c^2} \right) & \\
& \approx w_{ce} \left( 1 - \frac{v_r^2}{2c^2} + \frac{v_r^2}{c^2} \right) &\text{ ~~~~~~~~ First order approximation of $\Gamma _r^{-1} = (1-\frac{v_r^2}{c^2})^{\frac{1}{2}}$} \\
& & \text{~~~~~~~~using Taylor series expansion of the form $(1+x)^a = 1 + ax + o(x^n)$}\\
& \approx w_{ce} \left( 1+ \frac{v_r^2}{2c^2} \right) & \text{this is of the $(1+x)^a$ form,}\\
& & \text{so we can do a reciprocal Taylor expansion,}\\
& & \text{by posing $x=-\frac{v_r^2}{c^2}$ \& $a = -\frac{1}{2}$}\\
&= w_{ce} \left( 1- \frac{v_r^2}{c^2} \right)^{-1/2} &\text{ ~~~~~~~~ by doing the reciprocal Taylor series expansion}
\end{align*}
\begin{equation}
\omega_{lc} = \omega_{ce} ~ \Gamma_r
\label{resonance_lc_annex}
\end{equation}

In fine, this gives us for $N$:
\begin{equation} \label{indexNlcannexe}
N=\frac{\Gamma_r^{-1} ~v_r}{cos \alpha ~ c ~ cos \theta} = \frac{\chi}{cos \theta}
\end{equation}

Using the dispersion relation (Equation \ref{relation_dispersion}) and the equation above, we can rewrite it:

\begin{alignat}{2}
\begin{split}
 AN^4+BN^2+C & = 0 \\
 \Longleftrightarrow \frac{A \chi^4}{\cos^4 \theta} + \frac{B \chi^2}{\cos^2 \theta} + C  & = 0 \\
\Longleftrightarrow A \chi^4+B \chi^2cos^2\theta + C cos^4 \theta &  = 0
 \label{dispersion2}
 \end{split}
\end{alignat}

Thus, using the definitions of Equation \ref{coefficientsABC}:
\begin{equation}
\begin{split}
A&=P cos^2 \theta + S sin^2 \theta \\
B&=sin^2 \theta (D^2-S^2) - SP (1+ cos^2 \theta)\\
C &= P(S^2-D^2)
\end{split}
\end{equation}
with the definitions of S, P and D given Equation \ref{coefficientsSPD} page \pageref{coefficientsSPD}, the dispersion relation can be expressed as a function of $\theta$ as the only unknown:

\begin{alignat}{2}
\begin{split}
  \Longleftrightarrow & A \chi^4+B \chi^2cos^2\theta + C cos^4 \theta = 0 \\
  \Longleftrightarrow &  P \chi^4 \cos^2 \theta + \chi^4 S \sin^2 \theta + \chi^2 (D^2 - S^2 ) \sin^2 \theta \cos^2 \theta - \chi^2 P S (1 + \cos^2 \theta ) \cos^2 \theta  + P (S^2 - D^2) \cos^4 \theta  = 0
  \end{split}
  \end{alignat}
  with $\sin^2 \theta = 1 - \cos^2 \theta$\\
  \begin{alignat}{2}
  \begin{split}
  \Longleftrightarrow & P \chi^4 \cos^2 \theta + \chi^4 S (1 - \cos^2 \theta)  + \chi^2 (D^2 - S^2) \cos^2 \theta (1 - \cos^2 \theta) - \chi^2 P S \cos^2 \theta - \chi^2 PS \cos^4 \theta + P (S^2 - D^2) \cos^4 \theta = 0 \\
   \Longleftrightarrow & \color{blue}{P \chi^4 \cos^2 \theta} \color{teal}{+\chi^4 S} \color{blue}{-\chi^4 S \cos^2 \theta} \color{blue}{+ \chi^2 ( D^2 - S^2) \cos^2 \theta} \color{purple}{- \chi^2 (D^2 - S^2) \cos^4 \theta} \color{blue}{-\chi^2 P S cos^2 \theta} \color{purple}{- \chi^2 P S \cos^4 \theta} \color{purple}{+ P (S^2 - D^2) \cos^4 \theta}  \color{black}{= 0}
\end{split}
\end{alignat}

Finally, Equation \ref{dispersion2} can be rewritten as follows:
\begin{alignat}{2}
\begin{split}
  \Longleftrightarrow &  \color{purple}{\left[P(S^2-D^2) - \chi^2 (D^2 - S^2 + PS)\right]cos^4 \theta} + \color{blue}{\left[\chi^4 (P-S) + \chi^2(D^2 - S^2 -PS)\right]cos^2 \theta} + \color{teal}{\chi^4 S} \color{black}{=0} \\ 
  \Longleftrightarrow & \color{purple}{a}~cos^4 \theta + \color{blue}{b}~cos^2 \theta + \color{teal}{c}  \color{black}{= 0}
\end{split}
\end{alignat}

Therefore, the solutions of this equation are given by:
\begin{equation} \label{thetaLCannexe}
cos^2 \theta=\frac{-b\pm \sqrt{b^2-4ac}}{2a}
\end{equation}
with:
\begin{equation}
\begin{split}
 a = &P (S^2 - D^2) - \chi^2 (D^2 - S^2 +PS)\\
 b = & \chi^2 (P-S) + \chi^2 (D^2 - S^2 - PS)\\
 c = & \chi^4 S
\end{split}
\end{equation}

As for the solutions of $N$, the $\pm$ sign indicates there are two branches to this equation. As the loss-cone-driven ECM amplify waves only at frequency $f>f_{ce}$, and as only L-O and R-X modes tend to $N=1$ and therefore capable of propagating in the vacuum outside of the plasma, our only two solutions are:
\begin{itemize}
    \item[$+$]: LO
    \item[$-$]: RX
\end{itemize}

Calculation of the beaming angle in L-O or R-X mode, is available from Version 1.3.0 \citep{louis_2023_expresV130}, the version used in this article.

One should note that for ECM-loss-cone type simulations that would be done for an observer in the source of radio emissions, at a frequency between $f_{ce} < f <f_{UH}$, ExPRES would also be able to simulate emissions on the Z mode.

\newpage

\section{ExPRES simulations}
\label{appendix:expres_simulations}

Figure \ref{envelopesA} displays representative examples of simulated R-X emission envelopes with ExPRES, for loss-cone-driven ECM with several characteristic energies and for shell-driven ECM. Panels (a) and (b) are identical to Figure \ref{envelopes}.

\begin{figure}[H]
	\begin{center}
	\centerline{\includegraphics[width=1.0\linewidth]{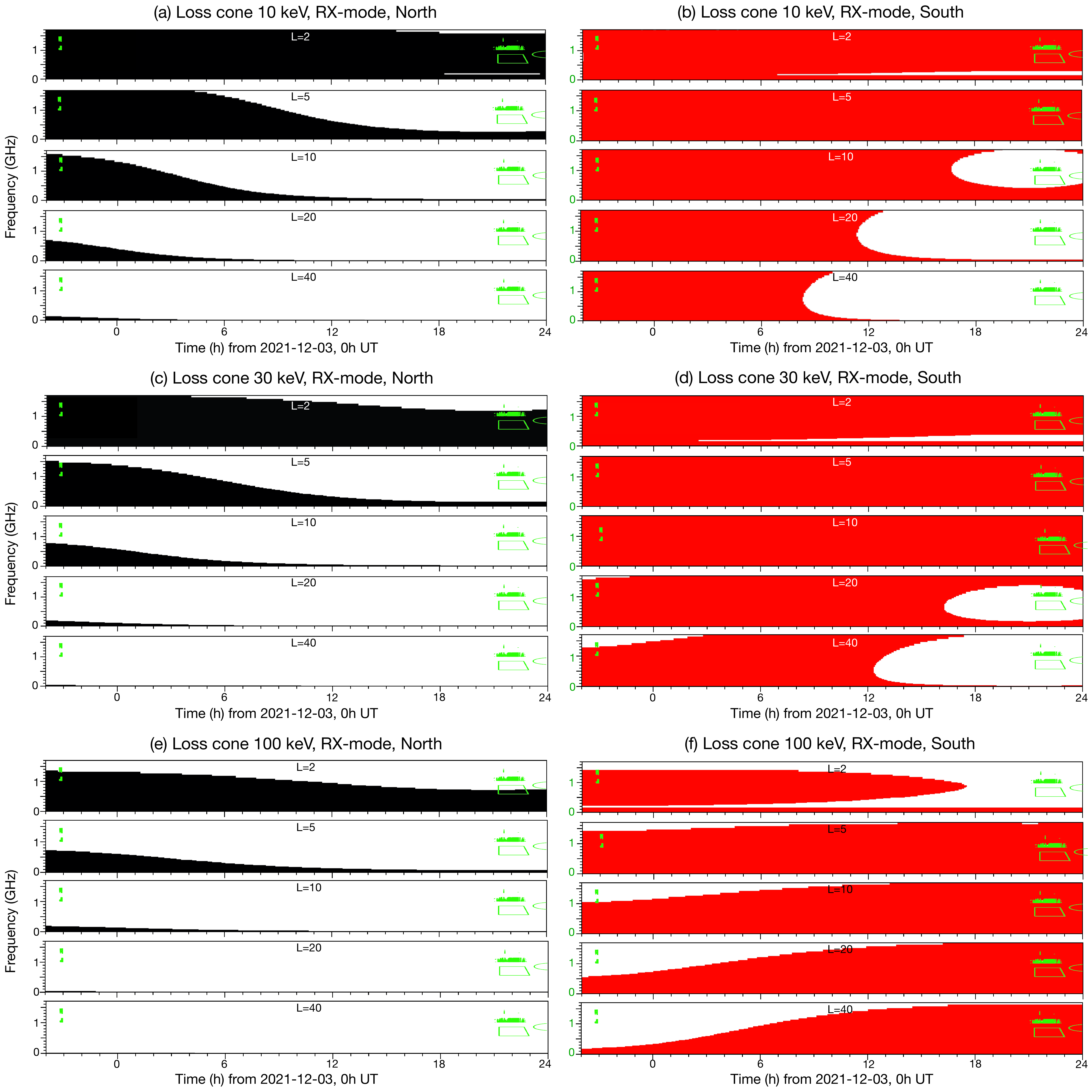}}
\caption{A few representative examples of simulated emission envelopes with ExPRES. R-X mode is emitted from the Northern (a,c,e) or Southern (b,d,f) hemisphere, by loss-cone-driven ECM with characteristic energy 10 keV (a,b), 30 keV (c,d), and 100 keV (e,f). Five dipolar magnetic shells (L=2, 5, 10, 20, 40) are simulated in each case, with active radiosources along all field lines at the corresponding shell (actually every degree of longitude \pz{-- small t-f gaps due to this discretization have been interpolated}). An Earth-based observer detects RH (black) or LH (red) polarized emission depending on the hemisphere of origin and emission mode. On all panels, bursts were detected by FAST in the green-shaded areas, while the green contours refer to uGMRT detections.}
\label{envelopesA}
\end{center}
\end{figure}

\setcounter{figure}{0}

\begin{figure}[H]
	\begin{center}
	\centerline{\includegraphics[width=1.0\linewidth]{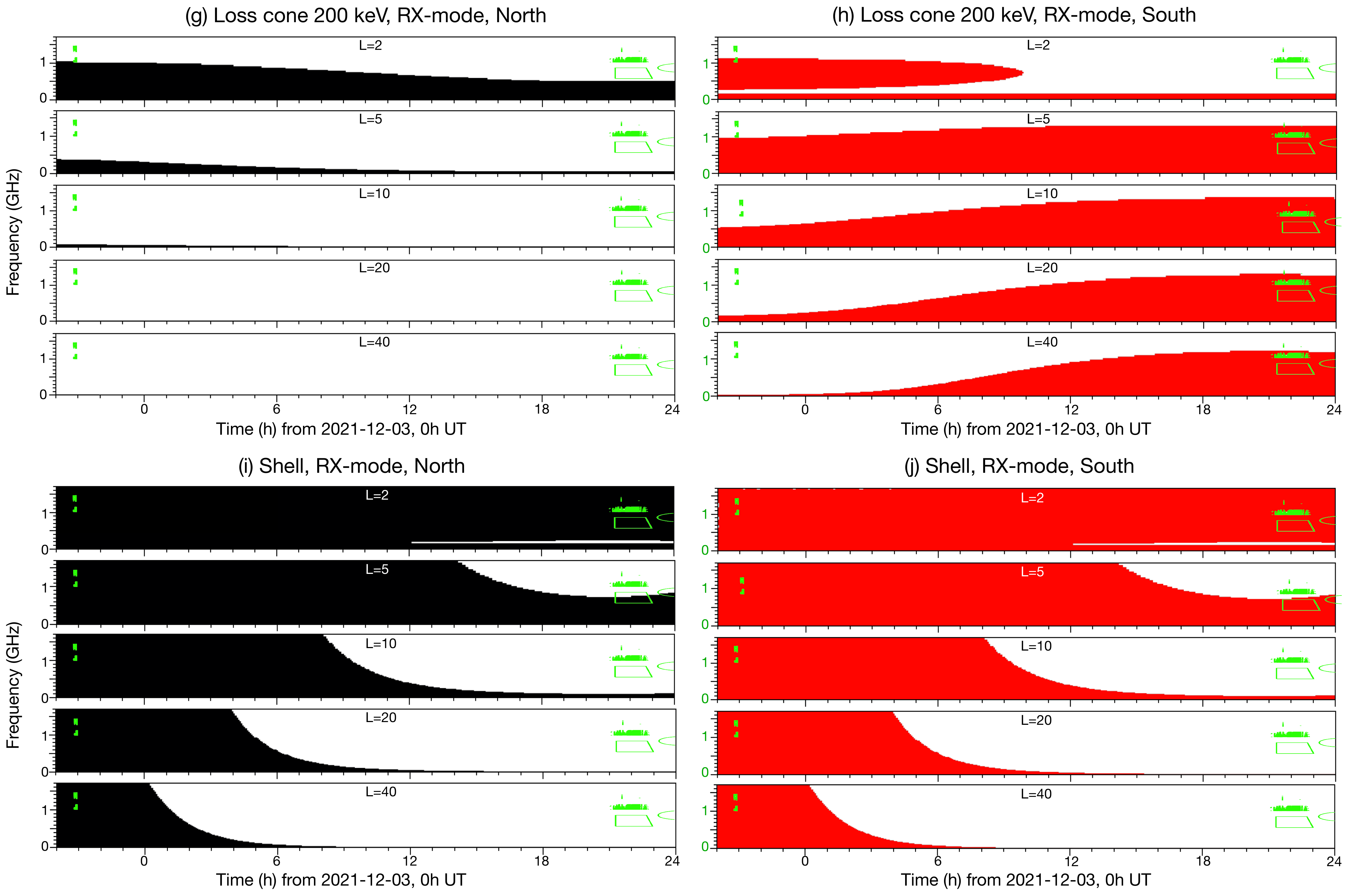}}
\caption{(continued), for loss-cone-driven ECM with characteristic energy 200 keV (g,h), and for shell-driven ECM (i,j).}
\end{center}
\end{figure}

\end{appendix}

\end{document}